\documentclass[fleqn,usenatbib,useAMS]{mnras}
\usepackage{newtxtext,newtxmath}

\usepackage[T1]{fontenc}
\usepackage{ae,aecompl}
\usepackage{siunitx}
\let\VANthebibliography\thebibliography
\def\thebibliography{\DeclareRobustCommand{\VAN}[3]{##3}\VANthebibliography}

\usepackage{graphicx}	
\usepackage{amsmath}	
\usepackage{amssymb}	
\usepackage{booktabs}
\usepackage{nccmath}
\usepackage{booktabs}
\usepackage{threeparttable}
\usepackage{caption}

\def\msun{{\rm M}_{\sun}}

\title[Lensing \& X-ray mapping of MS\,0451-03]{The distribution of dark matter and gas spanning six megaparsecs around the post-merger galaxy cluster MS\,0451$\mathbf{-}$03}

\author[Tam et al.\ 2020]{Sut-Ieng\ Tam$^{1}$\thanks{E-mail: sut-ieng.tam@durham.ac.uk},
Mathilde\ Jauzac$^{2,1,3}$,
Richard\ Massey$^{2}$,
David\ Harvey$^{4}$,
\newauthor
Dominique\ Eckert$^{5,6}$,
Harald\ Ebeling$^{7}$,
Richard\ S.\ Ellis$^{8}$,
Vittorio\ Ghirardini$^{5}$,
\newauthor
Baptiste\ Klein$^{9,10}$,
Jean-Paul\ Kneib$^{11}$,
David\ Lagattuta$^{2}$,
Priyamvada\ Natarajan$^{12}$,
\newauthor
Andrew\ Robertson$^{1}$ 
and Graham\ P.\ Smith$^{13}$
\\
\\
\\
$^{1}$Institute for Computational Cosmology, Durham University, South Road, Durham DH1 3LE, UK\\
$^{2}$Centre for Extragalactic Astronomy, Durham University, South Road, Durham DH1 3LE, UK\\
$^{3}$Astrophysics and Cosmology Research Unit, School of Mathematical Sciences, University of KwaZulu-Natal, Durban 4041, South Africa\\
$^{4}$Instituut-Lorentz for Theoretical Physics, Universiteit Leiden, Niels Bohrweg 2, Leiden, The Netherlands\\
$^{5}$Max-Planck Institut f\"ur Extraterrestrische Physik, Giessenbachstrasse 1, 85748 Garching, Germany\\
$^{6}$Astronomy Department, University of Geneva, 16 ch.\ d'Ecogia, CH-1290 Versoix, Switzerland\\
$^{7}$Institute for Astronomy, University of Hawaii, 2680 Woodlawn Drive, Honolulu, HI 96822, USA\\
$^{8}$Department of Physics and Astronomy, University College London, Gower Street, London WC1E 6BT, UK\\
$^{9}$Universit\'e de Toulouse, UPS-OMP, IRAP, 14 Avenue E. Belin, Toulouse F-31400, France\\
$^{10}$CNRS, IRAP/UMR 5277, Toulouse, 14 Avenue E. Belin, Toulouse F-31400, France\\
$^{11}$Laboratoire d'Astrophysique, Ecole Polytechnique F\'ed\'erale de Lausanne (EPFL), Observatoire de Sauverny, CH-1290 Versoix, Switzerland\\
$^{12}$Department of Astronomy, Yale University, 260 Whitney Avenue, New Haven, CT 06511, USA\\
$^{13}$School of Physics and Astronomy, University of Birmingham, Birmingham B15 2TT, UK
}

\pubyear{2019}
\begin{document}
\label{firstpage}
\pagerange{\pageref{firstpage}--\pageref{lastpage}}
\maketitle

\begin{abstract}
Using the largest mosaic of \emph{Hubble Space Telescope} images around a galaxy cluster, we map the distribution of dark matter throughout a $\sim$$6\times6$\,Mpc$^2$ area centred on the cluster MS\,0451$-$03 ($z=0.54$, $M_{200}=1.65\times10^{15}\,\msun$). Our joint strong- and weak-lensing analysis shows 
three possible filaments extending from the cluster, encompassing six group-scale substructures. The dark-matter distribution in the cluster core is elongated, consists of two distinct components, and is characterized by a concentration parameter of $c_{200}=3.79\pm0.36$.
By contrast, \textit{XMM-Newton} observations show the gas distribution to be more spherical, with excess entropy near the core, and a lower concentration of $c_{200}=2.35^{+0.89}_{-0.70}$ (assuming hydrostatic equilibrium).
Such a configuration is predicted in simulations of major mergers $2$--$7$\,Gyr after the first core passage, when the two dark-matter halos approach second turnaround, and before their gas has relaxed.
This post-merger scenario finds further support in optical spectroscopy of the cluster's member galaxies, which shows that star formation was abruptly quenched $5$\,Gyr ago. 
MS\,0451$-$03 will be an ideal target for future studies of the growth of structure along filaments, star-formation processes after a major merger, and the late-stage evolution of cluster collisions.

\end{abstract}


\section{Introduction}

The standard $\Lambda$CDM (Cold Dark Matter) model of cosmology suggests that large-scale structure formed hierarchically via a series of mergers with smaller halos and accretion of surrounding matter \citep{1978MNRAS.183..341W,Springel2005SimulationsOT,Schaye:2014tpa}. Reaching total masses of several $10^{15}\,\msun$, galaxy clusters are the largest and rarest structures resulting from this hierarchical formation process. Since their properties depend on the growth of structure (from the seeds provided by primordial density fluctuations, through gravitational collapses, to accretion of matter funnelled onto them along filaments), clusters are ideally suited to test cosmological models \citep[e.g.][]{1993ApJ...407L..49B,2005MNRAS.362.1301M,2010ApJ...708..645R,deHaan:2016qvy,Jauzac:2016tjc,Schwinn2017}.

Approximately 80\% of a cluster's mass consists of dark matter. Although this component is invisible, the total mass along a line of sight can be mapped through measurements of the deflection of light from background objects by gravitational lensing, a process that is independent of the physical or dynamical state of the lensing matter \citep[see reviews by e.g.][]{mrev,knrev,hrev,trev,kirev,srev}. The distinctive signatures of strong gravitational lensing (multiple images or giant arcs) probe the mass distribution in the inner region of clusters, while weak gravitational lensing provides constraints on the cluster environment on larger scales \citep[][]{2007ApJ...667..176G,2007Natur.445..286M,2012ApJ...748...56S,2013ApJ...762L..30Z}. 
Combining strong- and weak-lensing analyses can thus constrain the mass distribution across the entire cluster \citep{2005ApJ...619L.143B,2008ApJ...681..187B,jauzac18b}.

Wide-field observations of weak lensing with ground-based telescopes have successfully measured clusters' properties, including their mass \citep[e.g.][]{Umetsu:2014vna,2016MNRAS.461.3794O,2018PASJ...70S..28M,McClintock:2018bxh,Miyatake:2018lpb,2020ApJ...890..148U,Rehmann:2018nis,2019arXiv191204414H} and halo shape \citep[e.g.][]{Evans:2008mp,2010MNRAS.405.2215O,Clampitt:2015wea,vanUitert:2016guv,Shin:2017rch,Umetsu:2018ypz,Chiu:2018gok}. 
The latter is of particular interest since it reflects the nature of dark matter \citep[specifically whether dark matter is collisionless;][]{2019MNRAS.488.3646R}. On larger scales, the mass distribution's shape is governed by accretion of matter from the surroundings. As substructures are accreted onto clusters along filaments \citep[][]{2012MNRAS.426.2046A,1996Natur.380..603B,1996ApJ...465....2Y,2007ApJ...655L...5A}, cluster mass halos tend to align with the directions of infall \citep[e.g.][]{1992ApJ...399..405W,Jing:2002np}.
Direct detection of the mass in filamentary large-scale structures (LSS) through gravitational lensing is, however, extremely challenging with ground-based observations because of the filaments' low mass and the low density of resolved galaxies behind them \citep[e.g.][]{2006A&A...451..395C,Kaiser:1998ja,2002ApJ...568..141G,2004A&A...422..407G,2012Natur.487..202D,2016A&A...590A..69M}.

The higher angular resolution afforded by space-based imaging increases the signal-to-noise ratio (S/N) of lensing measurements. However, efforts to exploit this advantage are currently limited by the small field of view of the \emph{Hubble Space Telescope} (\emph{HST}). In the next decade, high-resolution observations from space over a much wider field of view will become possible with \emph{Euclid} \citep{2011arXiv1110.3193L}, \emph{Nancy Grace Roman Space Telescope} \citep[][]{2013arXiv1305.5422S}, and the balloon-borne telescope \emph{SuperBIT} \citep{2016arXiv160802502R,2018SPIE10702E..0RR}. It is thus timely to hone analysis methods that will exploit this new era of wide-field, high-resolution lensing data. 

Multi-wavelength data, including X-ray spectral imaging of the intra-cluster medium (ICM), are crucial to our understanding of the dynamics in clusters.
Since dark matter and baryons interact differently during a merger, a combined study of the distributions of dark matter and ICM provides insights into clusters' {evolutionary} history \cite[e.g.][]{2006ApJ...652..937B,2011MNRAS.417..333M,2015ApJ...812..153O,2015MNRAS.446.4132J,Molnar:2017ynu}. 
Furthermore, X-ray analyses usually assume that the ICM is in hydrostatic equilibrium (HSE) and spherically symmetric. Therefore, a comparison between X-ray and lensing mass measurements can be used to test the validity of the HSE assumption.

In this paper, we use a wide-field \emph{HST}/ACS imaging mosaic and \emph{XMM-Newton} observations to conduct a combined strong- and weak-lensing and X-ray analysis of the galaxy cluster MS\,0451$-$03 ($z=0.54$; hereafter MS\,0451), also known as MACS\,J0454.1$-$0300
\citep{2001ApJ...553..668E,2007ApJ...661L..33E} and the most X-ray luminous cluster in the Extended Medium Sensitivity Survey \citep[EMSS;][]{1990ApJS...72..567G}. MS\,0451 was extensively studied previously at optical wavelengths \citep[][]{1999A&AS..136..117L,2007ApJ...665.1067M,2007ApJ...671.1503M,2013AJ....145...77J,2015A&A...581A..31S, 2016A&A...590A..69M}, in X-rays \citep[]{2002ApJ...573L..91M,2003ApJ...598..190D,2005ApJ...624..606J,2018ApJS..235...29J}, and via the Sunyaev-Zel'dovich (SZ) effect \citep[]{2005ApJ...625..108D,2019ApJ...880...45S}. 
Strong gravitational lensing analyses have built a model of the cluster core \citep[][]{2004MNRAS.352..759B,2010A&A...509A..54B,2011MNRAS.410.1939Z,mackenzie2014,2020arXiv200610700J}, and a ground-based weak lensing analysis detected a possible filamentary structure \citep[]{2016A&A...590A..69M}.
In 2004, MS\,0451 was extensively observed with \emph{HST} over a large area, providing the community with the largest \emph{HST} mosaic centered on a galaxy cluster to date. In this paper, we exploit these wide \emph{HST} observations, combining strong and weak gravitational lensing to map the mass distribution out to a projected radius of $\sim$$3$\,Mpc.

This paper is organised as follows. The multiwavelength observations of MS\,0451 upon which we base our analysis are summarised in Section\,2. Our methods for measuring gravitational lensing and reconstructing the distribution of mass are described in Section\,3, while our X-ray data analysis is presented in Section\,4. Our measurements of the main cluster halo and surrounding large-scale structures are the subject of Section\,5. We discuss the cluster's dynamical state in Section~\ref{sec:dynamical}, before presenting our conclusions in Section~\ref{sec:concs}.

Throughout this paper, we adopt a $\Lambda$CDM cosmology with $\Omega_m=0.27$, $\Omega_{\Lambda}=0.73$, and $H_0=70$ km s$^{-1}$ Mpc$^{-1} $; 1$\arcsec$ corresponds to 6.49\,kpc at the redshift of the cluster. All magnitudes are quoted in the AB system.

\section{Observations}
\label{sec:obs}

\subsection{\emph{Hubble Space Telescope} observations}
\subsubsection{HST/\emph{ACS}}
A mosaic of 41 high-resolution images spanning $\sim$ $20\arcmin\times 20\arcmin$ around MS\,0451 was obtained with the \emph{Advanced Camera for Surveys} onboard \emph{HST} \cite[ACS;][]{1996SPIE.2807..184F} between January 19 and February 3, 2004 (GO-9836, PI: R.\ Ellis), in the F814W pass-band, with an exposure time of 2\,ks per pointing (single-orbit depth). We reduced the data with the \textsc{pyHST} software package\footnote{\url{https://github.com/davidharvey1986/pyHST}} which corrects Charge Transfer Inefficiency using \textsc{arctic} \citep{2014MNRAS.439..887M}, removes detector bias and applies flat-field corrections using \textsc{calacs} \citep{isr1805}, and finally stacks the dithered images using \textsc{astrodrizzle} \citep{isr1702}.
We use these high-resolution images to measure the effect of weak gravitational lensing on the shapes of background galaxies. 
\subsubsection{HST/\emph{WFC3}}
A $2\arcmin\times2\arcmin$ region in the cluster core was imaged with the \emph{Wide Field Camera 3} onboard \emph{HST} \citep[WFC3;][]{kimble2008} on January 13, 2010 (GO-11591, PI: J.-P.\ Kneib). We use observations in the F110W and F160W passbands, with exposure times of 17\,912\,s and 17\,863\,s respectively, for the strong-lensing analysis.

\subsection{Ground-based Observations}
\label{sec:obs_gb}
\subsubsection{Imaging Data}
\label{sec:obs_gb_imaging}
Multicolour imaging in the $B$,$V$,$R_{c}$,$I_{c}$ and $z'$ passbands was obtained with the 8.3\,m Subaru telescope's wide-field Suprime-Cam camera for 1440\,s, 2160\,s, 3240\,s, 1800\,s, and 1620\,s, respectively. 
Observations were performed on December 21, 2006 ($z'$), December 11, 2001 ($R_{c}$,$I_{c}$), and January 23, 2009 ($B$,$V$). Near-UV imaging in the $u^{*}$ passband was obtained with the 3.6\,m CFHT's MegaPrime camera for 6162\,s on November 27, 2006 (ID: 06BH34, PI: H.\ Ebeling). Near-infrared observations in the $J$ and $K_S$ passbands were performed with CFHT's Wide-field InfraRed Camera (WIRCam) on November 8, 2008 and October 25, 2007, respectively (ID: 08BH63, 07BH98, PI: C.-J.\ Ma). All observations were dithered to facilitate the removal of cosmic rays and minimise the impact of pixel defects and chip gaps; all data were reduced using standard procedures \citep{Donovan}.

We use these data to measure photometric redshifts and thereby identify galaxies within, in front of, or behind the cluster. In order to allow a robust estimate of the spectral energy distribution (SED) to be obtained for all objects within the field of view, data from different passbands were seeing-matched using the technique described in \cite{2008MNRAS.389.1240K}. The object catalogue was then created with the \textsc{SExtractor} photometry package \citep[][]{1996A&AS..117..393B} in `dual-image mode', with the $R_C$-band image as the reference detection image. Photometric redshifts for galaxies with magnitude ${R_{C}}<24$ were subsequently computed using the adaptive SED-fitting code \textsc{Le Phare} \citep[][]{1999MNRAS.310..540A, 2009ApJ...690.1236I}. For more details of this procedure, see \cite{2008ApJ...684..160M}.

\subsubsection{Spectroscopic observations}

MS\,0451 was also observed with the Multi-Unit Spectroscopic Explorer \citep[MUSE;][]{bacon2010} at the VLT on January 10-11, 2016 (ID: 096.A-0105(A), PI: J.-P.\ Kneib), in WFM-NOAO-N mode and good seeing of approximately 0.8\arcsec. The MUSE observations consist of two pointings of three exposures, slightly shifted to account for systematic variations in the detector response, and cover a field of view of $\sim$ $2.2\,\rm{arcmin}^2$. These data are used for the strong-lensing analysis. They were reduced using version~1.6.4 of the MUSE standard pipeline \citep{Weilbacher2012,Weilbacher2014}, which applies bias and illumination corrections; performs geometrical, astrometric and flux calibrations; and then combines the individual exposures for each pointing into a single data cube. The sky residuals within each data cube were subtracted using the Zurich Atmosphere Purge algorithm \citep{soto2016}, which masks sources identified by \textsc{sextractor} \citep{1996A&AS..117..393B}, and then uses principal component analysis to model the sky background.

The spectroscopic redshifts of galaxies used in this work were compiled from the literature and complemented by redshifts measured by us, based on spectroscopic observations obtained in September 2004 with Gemini-North/GMOS on Mauna Kea. The latter used a 1\,arcsec slit, the 800\,l/mm grating, and a spectral range from approximately 4200 to 7000\,\AA. The resulting data were reduced using standard IRAF procedures.

\subsection{\emph{XMM-Newton} X-ray Observations}
\label{sec:xrayobs}

MS\,0451 was observed by \emph{XMM-Newton} (observation ID: 0205670101, PI: D. Worrall) on September 16-17, 2004 for a total of 44\,ks. We reduced the \emph{XMM-Newton}/EPIC data using the {\textsc XMMSAS} v16.1 software package and a pipeline developed in the framework of the \emph{XMM-Newton Cluster Outskirts Project} \citep[X-COP,][]{Eckert2017}. 
After performing the standard data reduction steps to extract calibrated event files, we used the XMMSAS tools \textsc{mos-filter} and \textsc{pn-filter} to automatically define good-time intervals (unaffected by soft proton flares) of 24\,ks (MOS1), 24\,ks (MOS2), and 19\,ks (PN). 
For more details of this procedure, see \citet[][Sect.~2 and Fig.~1]{Ghirardini2019}.
These data are used to measure the properties of the baryonic ICM.

An independent, $\sim50$ks  \emph{Chandra} observation  provides a high-resolution X-ray view of the cluster core but is not used by us here, since the covered area does not match the extended \emph{HST} mosaic that is the focus of this paper.

\section{Method: Gravitational Lensing Analysis }
\label{sec:wl}

\subsection{Weak-lensing theory}
\label{sec:wlth}
Weak gravitational lensing is caused by gravitational fields (here created by the massive cluster MS\,0451) that deflect  light rays emitted by background galaxies, distorting their apparent size and shape. Projecting the cluster's 3D mass distribution along the line of sight yields a 2D surface density $\Sigma(\boldsymbol{R})$, where ${\boldsymbol{R}}=(x,y)$ is a position in the plane of the sky. The cluster's gravitational field causes an isotropic magnification of background galaxies by a factor known as `convergence'
\begin{equation}
\kappa(\boldsymbol{R})=\Sigma(\boldsymbol{R})/\Sigma_c,
\label{eq:kappa}
\end{equation}
where
\begin{equation}
\Sigma_c=(c^2 D_s)/(4\pi GD_lD_{ls})
\end{equation}
is the critical surface mass density for lensing, and $D_l$, $D_s$, $D_{ls}$ are the angular diameter distances from the observer to the lens, from the observer to the source, and from the lens to the source, respectively. 

While $\kappa$ is difficult to observe, a related quantity is more readily measurable. $\Sigma(\boldsymbol{R})$ also induces a shear distortion
\begin{equation}
\boldsymbol{\gamma}=\gamma_1+i\gamma_2=\lvert\boldsymbol{\gamma}\rvert e^{2i\phi},
\label{eq:gamma_ini}
\end{equation}
whose real (imaginary) component is reflected in the apparent elongation of background galaxies along (at $45^\circ$ to) an arbitrarily oriented real axis.
A combination of shear and convergence known as `reduced shear' $\boldsymbol{g}\equiv\boldsymbol{\gamma}/(1-\kappa)$ can be measured from the ellipticities of background galaxies 
\begin{equation}
\boldsymbol{\epsilon}=\boldsymbol{\epsilon}_{\text{int}}+G \boldsymbol{g} 
~, \label{eq:etogamma}
\end{equation}
where $\boldsymbol{\epsilon}_{\text{int}}$ is a galaxy's intrinsic ellipticity, and $G$ is a `shear susceptibility' factor (see Section~\ref{sec:RRG}).
Although the unknown intrinsic shapes of galaxies are a dominant source of noise, no bias is introduced if the lensed background galaxies are randomly oriented, $\langle\boldsymbol{\epsilon}_{\text{int}}\rangle=0$.

\subsection{Weak-lensing measurements}
\subsubsection{Source detection}

We identify galaxies in the \emph{HST}/ACS imaging mosaic from source properties determined with the \textsc{SExtractor} photometry package \citep[][]{1996A&AS..117..393B}. To maximise sensitivity to distant (small and faint) galaxies that contain most of the lensing signal, we adopt the `Hot--Cold' technique \citep[]{2004ApJS..152..163R}, i.e., we first create a source catalogue using a `cold' configuration, designed to detect only the brightest objects, and then a second one with a `hot' configuration, optimised to detect faint objects. We then merge the two catalogues, removing any `hot' sources that already have `cold' detections. We also remove any objects located near saturated stars or saturated pixels, using polygonal mask regions defined by hand.
Using \textsc{SExtractor} parameters, we assign each source a signal-to-noise ratio of detection, defined as  $\text{S/N}\equiv\text{FLUX\_AUTO}/\text{FLUXERR\_AUTO}$,
and classify objects as galaxies, stars or spurious features based on their overall brightness ($\text{MAG\_AUTO}$) and peak surface brightness ($\text{MU\_MAX}$). The resulting \emph{HST}/ACS catalogue contains 57,281 galaxies.

\subsubsection{Weak-lensing shape measurements}
\label{sec:RRG}

We measure the shapes of galaxies detected in the \emph{HST}/ACS images with the {\sc pyRRG} \citep{2019arXiv191106333H} implementation of the shear measurement method of \citet[][hereafter RRG]{2000ApJ...536...79R}  
which was designed to correct the small, diffraction-limited point spread function (PSF) of space-based observatories and has been calibrated on simulated data containing a known shear \citep{2007ApJS..172..219L}.

We model the PSF from stellar images in each exposure. The size and ellipticity of the PSF varies over time, as thermal expansion and contraction of the telescope change the distance between the primary and secondary mirrors and hence the focus. We determine the effective focus of each exposure by comparing the ellipticity of observed stars with models created with the \textsc{tinytim} ray-tracing software \citep[\textsc{tinytim},][]{2007ApJS..172..203R} and stacked to mimic the drizzling of multiple exposures. We interpolate the moments of the net PSF's shape using a polynomial fitting function.

We then determine the shapes of galaxies to extract the needed weak-lensing information. We measure the second and fourth order moments of each galaxy,
\begin{equation}
I_{ij}=\frac{\sum \omega(x,y) x_ix_j I(x,y)}{\sum \omega(x,y) I(x,y)},
\end{equation}
\begin{equation}
I_{ijkl}=\frac{\sum \omega(x,y) x_i x_j x_k x_l I(x,y)}{\sum \omega(x,y) I(x,y)},
\end{equation}
where $I$ is the intensity recorded in a pixel, $\omega$ is a Gaussian weight function included to suppress noise, and the sum is taken over all pixels. Each measured moment is corrected for convolution with the telescope's PSF. 
We subsequently calculate each galaxy's size
\begin{equation}
d=\sqrt{\frac{(I_{xx}+I_{yy})}{2}}
\end{equation}
and ellipticity $\epsilon\equiv\sqrt{\epsilon_1^2+\epsilon_2^2}$, using the definitions

\begin{equation}
\epsilon_1=\frac{I_{xx}-I_{yy}}{I_{xx}+I_{yy}}\ , and
\end{equation}
\begin{equation}
\epsilon_2=\frac{2I_{xy}}{I_{xx}+I_{yy}}\ .
\end{equation}

Applying equation~\eqref{eq:etogamma}, we finally obtain a shear estimator
\begin{equation}
\tilde{\boldsymbol{g}}=C\frac{\boldsymbol{\epsilon}}{G}
\label{eq:sheartilde}
\end{equation}
from each galaxy. Here the `shear susceptibility factor' $G$ is measured from the global distribution of $\epsilon$ and fourth-order moments \citep[]{2000ApJ...536...79R}. The calibration factor $C$, defined by $1/C=0.86^{+0.07}_{-0.05}$, is empirically measured from mock \emph{HST} images in the same band and of the same depth \citep[]{2007ApJS..172..219L}.
We note that shapes of very small or faint galaxies are difficult to measure and may be biased. As in the calibration tests, we exclude galaxies with size $d<0.11\,\arcsec$, S/N\,$<4.5$, or unphysical values of $\epsilon>1$
(which can arise after PSF correction in the presence of noise; for a discussion of this effect, see \citealt{2012MNRAS.426.3369J}).

\subsubsection{Identification of galaxies \emph{behind} the cluster}

The \emph{HST}/\emph{ACS} galaxy catalogue contains not only background galaxies but also foreground galaxies and cluster members that are not gravitationally lensed by the cluster. These unlensed galaxies dilute the shear signal. We use multicolour ground-based imaging to identify and eliminate them from our catalogue.

Robust photometric redshifts (see section~\ref{sec:obs_gb_imaging}) can be assigned to the 13\% of galaxies in the \emph{HST}/ACS catalogue that are brighter than ${R_{C}}=24$.
Based on these redshifts, we remove as likely cluster members all galaxies with $0.48 < z_\mathrm{phot} < 0.61$. For galaxies with spectroscopic redshifts we use the the more stringent criterion $0.522 < z_\mathrm{spec} < 0.566$ to eliminate cluster members.

\begin{figure}
\includegraphics[width=0.5\textwidth]{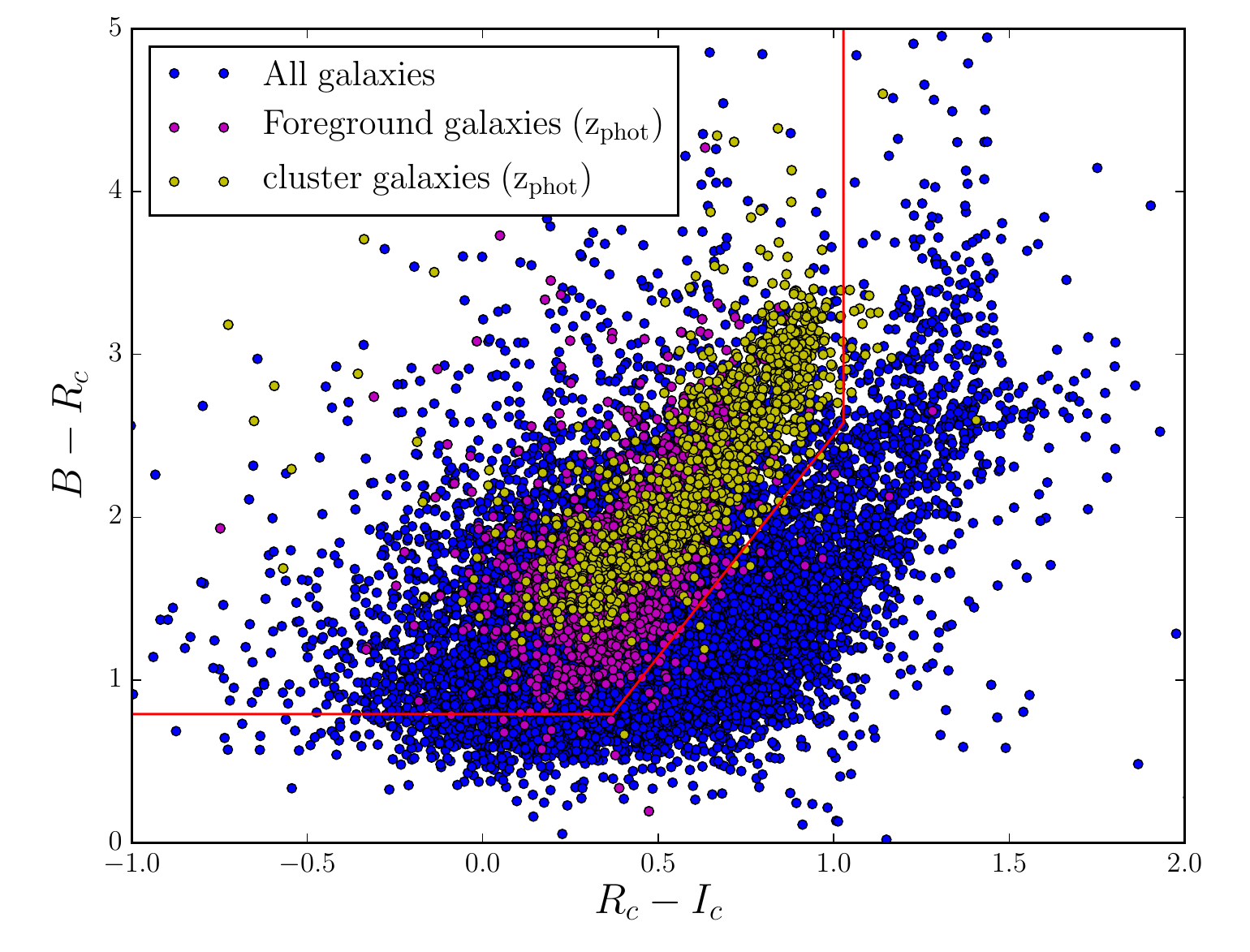}
\caption{Colour-colour diagram ($B-R_C$ vs $R_C-I_C$) for objects within the \emph{HST}/ACS mosaic of MS\,0451. Blue dots represent all objects; magenta and yellow dots are galaxies classified as foreground or cluster galaxies, respectively, based on photometric redshifts. The red solid lines delineate the $B$, $R_C$ and $I_C$ colour cuts applied to minimize contamination of the catalogue by unlensed objects.}
\label{fig:ccsel}
\end{figure}

\begin{figure}
\includegraphics[width=0.5\textwidth]{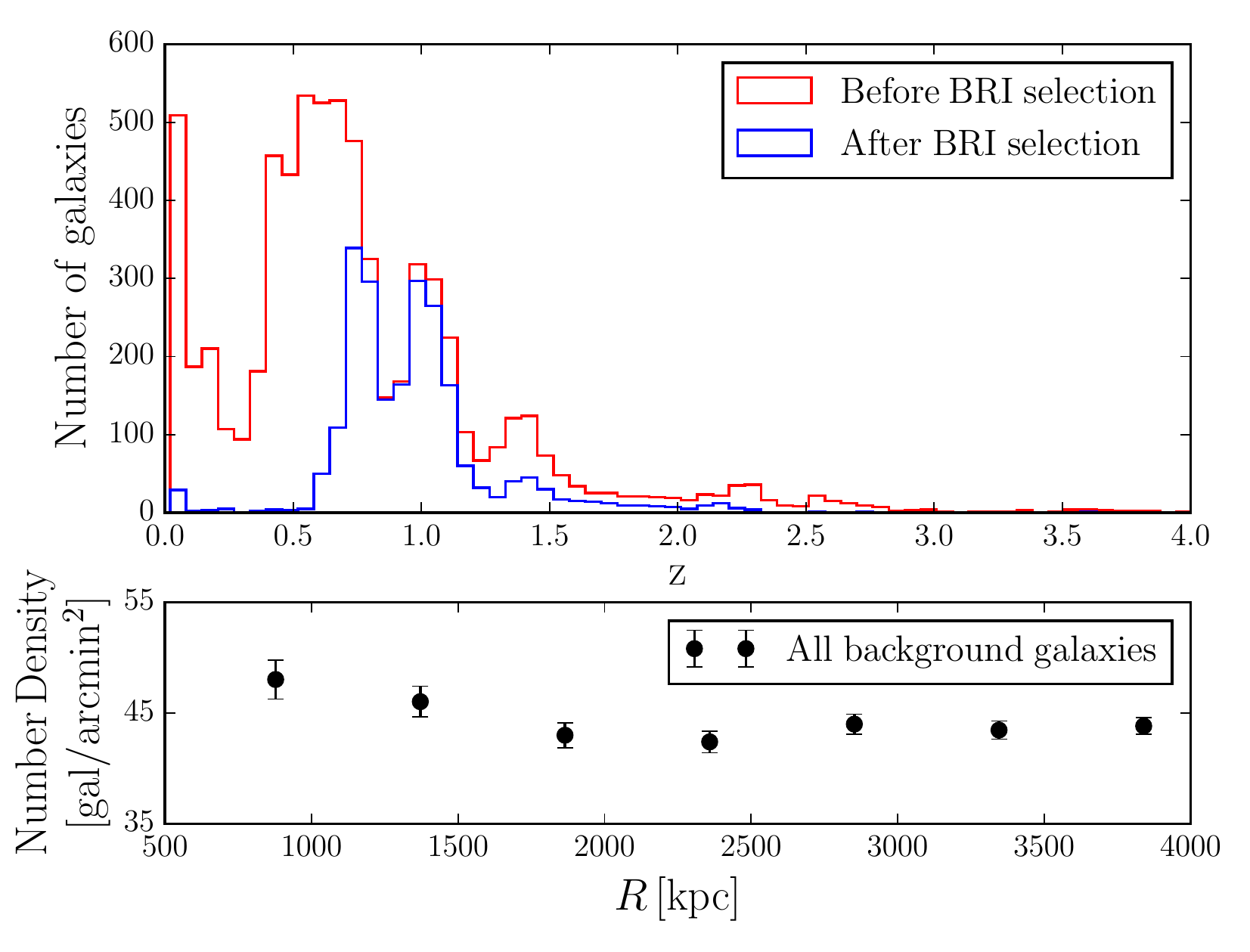}
\caption{Identification of background galaxies. \emph{Top:} Redshift distribution of all galaxies with spectroscopic or photometric redshifts (red histogram). The blue histogram shows the redshift distribution of galaxies classified as background sources based on $B$, $R_C$, $I_C$ colour-colour selection. \emph{Bottom:} Number density of {\it all} background galaxies in the final weak-lensing catalogue (including fainter galaxies without observed colours), as a function of their projected distance from the cluster center.}
\label{fig:zdistri}
\end{figure}

For an additional 16\% of galaxies, we obtain multicolour photometry in at least the $B$, $R_C$, and $I_C$ bands. After testing several criteria chosen elsewhere in the literature to identify foreground and cluster member galaxies (e.g., cuts in $B-R_C$ and $R_C-I_C$, or $B-V$ and $u-B$; see \citealt{2010MNRAS.405..257M,2018PASJ...70...30M,2012MNRAS.426.3369J}) we adopt cuts that retain only those galaxies with $(B-R_C)$$<$$0.79$, $(R_C-I_C)$$>$$1.03$, or $(B-R_C)$$<$$2.72(R_C-I_C)-0.216$ (Fig.~\ref{fig:ccsel}).
After these colour cuts are applied, the photometric redshift distribution of ${R_{C}}<24$ galaxies suggests a remaining contamination from foreground galaxies and cluster members of $\sim$4$\%$ (Fig.~\ref{fig:zdistri} top panel), which is smaller than our statistical error. We shall refer to the combined 30\% of galaxies with photometric information as the `bright sample'.

From the remaining 70\% of galaxies without ground-based photometric information, we next discard the 6\% of galaxies that are brighter than ${F814W}<24$. These are mainly foreground or cluster member galaxies: in the bright sample, 80\% of foreground galaxies and 89\% of cluster members have ${F814W}<24$, and their combined magnitude distribution peaks at ${F814W}\sim23$. 
For the remaining `faint sample' of galaxies, we assign nominal redshifts drawn at random from a distribution $N(z>0.54)\propto (e^{-z/z_0})^\beta$, with $\beta=1.8$ and median redshift $z_0=0.71$ \citep[]{1997MNRAS.287..833N,2009MNRAS.396..354G}. 

Our final weak-lensing catalogue (combining the `bright' and `faint' samples) contains 21,\,232 background galaxies, at a density of $44$\,galaxies\,arcmin$^{-2}$. 
Before cuts, the galaxy density shows an excess of $\sim35\,$galaxies\,arcmin$^{-2}$ within $1$\,Mpc of the cluster centre; our selection process removes this excess, leaving an approximately constant density throughout the field (Fig.~\ref{fig:zdistri} bottom panel), as expected for an uncorrelated population of background galaxies.

Of our final sample of background galaxies, 10\%, 11\% and 79\% are selected via cuts in redshift, colour and magnitude, respectively.

\subsection{Strong-lensing constraints}
\label{sec:sl}

For this analysis, we adopt the best-fit strong-lensing mass model from \cite{2020arXiv200610700J}. We here only give a summary of the strong-lensing mass model, and refer the reader to \cite{2020arXiv200610700J} for more details.
The cluster core is modeled using two cluster-scale halos and 144 galaxy-scale halos associated with cluster galaxies. All potentials are modeled using Pseudo-Isothermal Elliptical Mass Distributions \citep[PIEMDs;][]{kassiola1993,2005MNRAS.356..309L,eliasdottir2007} which are described by seven parameters: position ($x$,$y$), ellipticity $e$, position angle $\theta$, core radius $r_{\text{core}}$, truncation radius $r_{\text{cut}}$, and velocity dispersion $\sigma$.

The 2D surface mass density of each PIEMD is described by 
\begin{equation}
\label{eq:PIEMD}
\Sigma(R) = \frac{\sigma^{2}}{2G} \frac{r_{\text{cut}}}{r_{\text{cut}}-r_{\text{core}}} \left( \frac{1}{\sqrt[]{R^{2} + r_{\text{core}}^{2}}} - \frac{1}{\sqrt[]{R^{2} + r_{\text{cut}}^{2}}} \right),
\end{equation}
where the projected radius $R^{2} = x^{2}/(1+e_{\Sigma})^{2} + y^{2}/(1-e_{\Sigma})^{2}$ is defined by an ellipticity $e_{\Sigma}\equiv(a-b)/(a+b)$ with semi-major axis $a$ and semi-minor axis $b$ \citep{kassiola1993,1997MNRAS.287..833N}. Note that \textsc{Lenstool} reports ellipticity $e\equiv(a^2-b^2)/(a^2+b^2)$ and internally converts $e$ into $e_{\Sigma}$ during optimisation.
Best-fit parameters for the two cluster-scale components are listed in Table~\ref{tab:sl_model}.

Seven cluster galaxies acting as small-scale perturbers of some of the multiple images are independently modeled as individual PIEMDs. The rest of the cluster galaxy population is modeled using scaling relations; to limit the number of free parameters, positions, ellipticities, and position angles of all galaxies are fixed to the respective values of the observed stellar component. The galaxies' velocity dispersions are scaled from the observed stellar luminosity according to the \cite{faber1976} relation, which describes well the mass in early-type cluster galaxies \citep{2004ApJ...605..677W,2007NJPh....9..447J}.  

The strong-lensing mass model is constrained by 16 systems of multiple images (47 images in total). These include well known lensed objects such as a sub-millimeter arc at $z\sim 2.9$ \citep{2004MNRAS.352..759B}, five other sub-millimeter systems \citep{mackenzie2014}, a triply imaged galaxy \citep{takata2003}, and six new systems identified with VLT/MUSE, including a quintuple image at $z=6.7$ in the poorly constrained northern region. 
The latter has a redshift from VLT/XShooter observations and was previously studied by \cite{2016MNRAS.462L...6K}.
Five of these systems are spectroscopically confirmed, two of them newly identified by \cite{2020arXiv200610700J} using MUSE observations. The quintuple-image system in combination with the two new systems identified through MUSE observations in the northern region motivated the addition of a second cluster-scale halo in the strong-lensing mass model. Without this second large-scale halo, the geometry of the $z=6.7$ system cannot be recovered and the root-mean-square (rms) distance between the observed and predicted locations of the multiple images of other systems is unacceptably high at $>1.5$\arcsec. Two close groups of cluster galaxies were identified in this region. Adding a third cluster-scale mass halo did not significantly improve the model.

The resulting best-fit strong-lensing mass model has an rms separation of 0.6\arcsec. The best-fit parameters of the two main cluster halos are given in Table\,\ref{tab:sl_model}. Note that the coordinates of the halos are given in arcseconds relative to the cluster center, here the BCG ($\alpha=$73.545202, $\delta=-$3.014386).

\begin{table}
\centering
\begin{tabular}{ccc}
\hline
\hline
\textbf{Parameter} & \textbf{Main halo} & \textbf{Second halo} \\
\hline\\[-2mm]
R. A. & $-7.5^{+0.9}_{-1.2}$ & $22.3^{+3.1}_{-0.1}$ \\[1mm]
Dec. & $-2.6^{+0.6}_{-0.7}$ & $19.5^{+4.8}_{-0.1}$  \\[1mm]
$\sigma$ (km/s) & $1001^{+30}_{-25}$ & $810^{+210}_{-670}$ \\ [1mm]
$e$ & $0.63^{+0.04}_{-0.03}$ & $0.18^{+0.12}_{-0.06}$\\[1mm]
$\theta$ (deg) & $32.2 \pm 0.5$ & $147^{+9}_{-16}$ \\[1mm]
$r_{\text{core}}$ (kpc) & $120^{+10}_{-6}$ & $332^{+60}_{-30}$\\[1mm]
$r_{\text{cut}}$ (kpc) & [1000] & $680^{+200}_{-570}$ \\[2mm]
\hline
\hline
\end{tabular}
\caption{Best-fit parameters for the two cluster-scale halos included in the strong-lensing mass model of MS\,0451.
Coordinates are given in arcseconds relative to the location of the BCG ($\alpha=$73.545202, $\delta=-$3.014386). Since the truncation radius of the larger halo is far larger than the effective radius of the strong-lensing regime, it was frozen at 1$\,$Mpc. For more details see \citet{2020arXiv200610700J}.} 
\label{tab:sl_model}
\end{table}

\subsection{Lensing 2D mass map}
\label{sec:massrec}

We reconstruct the projected (2D) mass distribution using version~7.1 of 
\textsc{Lenstool}\footnote{Available at \url{https://projets.lam.fr/projects/lenstool/wiki}.} \citep{2007NJPh....9..447J,2009MNRAS.395.1319J,2012MNRAS.426.3369J}, whose performance has been quantified on mock \emph{HST} data by \cite{2020arXiv200610155T}. Specifically, we compute the mean mass map from 1700 Markov Chain Monte Carlo (MCMC) samples from the posterior. We find consistent but noisier results for the model parameters if we use the \citet{1993ApJ...404..441K} direct-inversion method (see Appendix A).

\subsubsection{Mass model}

\begin{figure}
\includegraphics[width=0.47\textwidth]{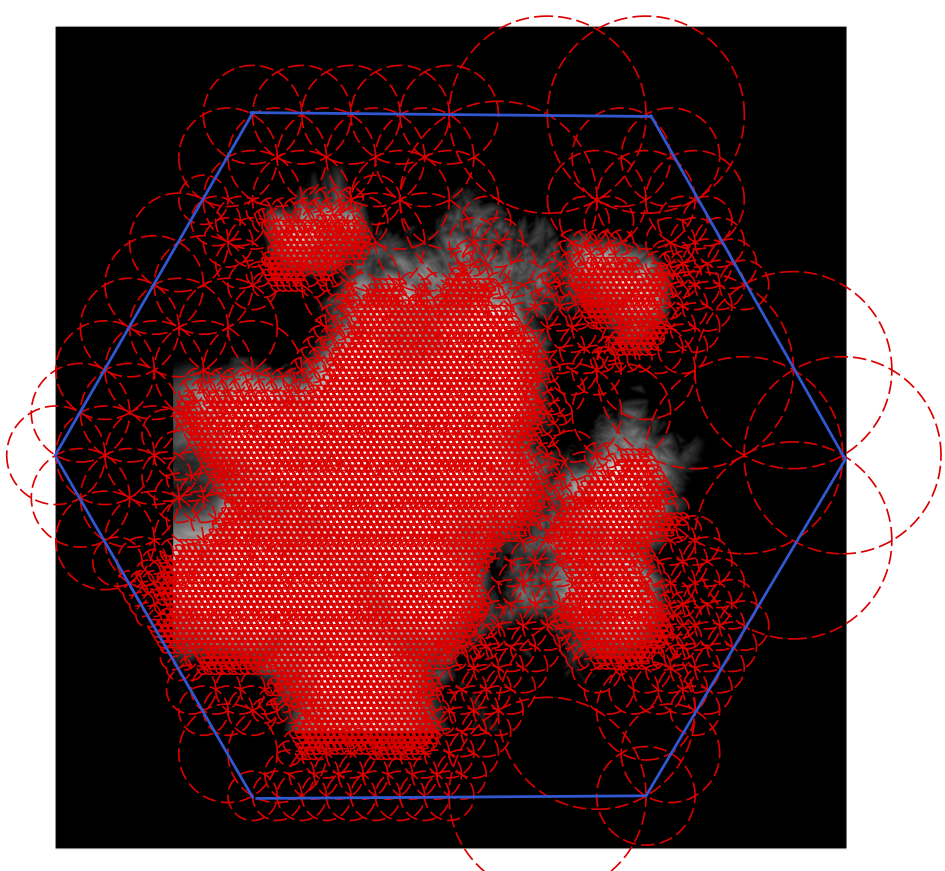}
\caption{The multi-scale grid that determines the maximum spatial resolution of the {\sc Lenstool} mass reconstruction. One RBF is placed at the centre of each circle, with core radius $r_\mathrm{c}$ equal to the radius of the circle, and a free mass normalisation. The grid is determined from (and shown overlaid upon) the cluster's $K$-band emission. The blue hexagon covers an area slightly larger than the {\em HST} field of view.}
\label{fig:grid}
\end{figure}

We set the mass distribution in the cluster core to our strong-lensing model (see Sect~\ref{sec:sl}), which consists of two cluster-scale halos separated by 237\,kpc and seven individually optimised galaxy-scale components.

To extend our analysis from the cluster center to $\sim$3\,Mpc, we add a total of 1277 galaxy-scale halos at the locations of cluster member galaxies, identified via spectroscopic and photometric redshifts over the entire field of view covered by the \emph{HST} mosaic. Each of these is modelled as a PIEMD potential with fixed parameters  $r_{\text{core}}=0.15\,\text{kpc}$ and $r_{\text{cut}}^*=58\,\text{kpc}$, and a velocity dispersion $\sigma$ that is scaled relative to an $m^*_{K}=18.7$ galaxy with $\sigma^*=163.1\,\text{km\,s}^{-1}$ using the \citet{faber1976} relation. Throughout the mosaic area outside the strong-lensing region, we then add a grid of masses, whose normalisation is allowed to vary and whose resolution is adapted to the local $K$-band luminosity. 
Following the procedure illustrated in Fig.~1 of \citet{2009MNRAS.395.1319J}, we create the multi-resolution grid by first drawing a large hexagon over the entire field of view, and then splitting it into six equilateral triangles. If a single pixel inside any of these triangles exceeds a predefined threshold in surface brightness, we split that triangle into four smaller triangles. This refinement continues for six levels of recursion, until the brightest parts of the cluster are sampled at the highest resolution, corresponding to a triangle side of 16\arcsec. Once this tessellation process is complete, we place a circular ($e=0$) PIEMD halo (Eq.~\ref{eq:PIEMD}) at the centre of each triangle, with a core radius $r_{\mathrm{c}}$ equal to the side length of the triangle, a truncation radius $r_{\mathrm{t}}=3r_{\mathrm{c}}$, and a velocity dispersion that is left free to vary. To avoid superseding the strong-lens model, we prevent the mass grid from extending into the multiple image region, defined as an ellipse aligned with the cluster core ($a=44\arcsec$, $b=34\arcsec$, $\theta=30^\circ$ counter-clockwise with respect to the East-West axis, centered on $\alpha$$=$$73.545202^\circ$, $\delta$$=$$-3.0143863^\circ$). We also exclude shear measurements from this region.
Our final grid (Figure~\ref{fig:grid}) for MS\,0451 model includes 5570 individual Pseudo-Isothermal Mass Distributions.

We optimise masses in the grid using the Bayesian \textsc{MassInf} algorithm and the Gibbs approach to maximise the likelihood
\begin{equation}
\label{eq:lenstoollikelihood}
    \mathcal{L}_{\gamma}=\frac{1}{Z_L}\rm{exp}\left(-\frac{\chi^2}{2}\right),
\end{equation}
where the goodness-of-fit statistic is
\begin{equation}
\label{eq:lenstoolchisquared}
\chi^2=\sum_{i=1}^{M}\sum_{j=1}^{2}\frac{\left({\gamma}_{j,i}-\gamma_{j,i}^\mathrm{model}\left(\boldsymbol{R}_i\right)\right)^2}{\sigma_{\gamma}^2}
\end{equation}
(following \citealt{2000A&A...353...41S})\footnote{Note that \textsc{Lenstool} takes inputs in the form of ellipticity $e=({a^2-b^2})/({a^2+b^2})$ instead of shear \citep{2014MNRAS.437.3969J}, so we use $\gamma^\mathrm{model}=2e^\mathrm{model}$.}, where $M$ is the number of background sources and the normalisation is
\begin{equation}
\label{eq:lenstoollikelihood_normalisation}
{Z_L}=\prod_{i=1}^{M}\sqrt{2\pi}\,\sigma_{\gamma i}~.
\end{equation}
Equations~\eqref{eq:lenstoolchisquared} and \eqref{eq:lenstoollikelihood_normalisation} involve the statistical uncertainty on each shear measurement, $\sigma_{\gamma}$. We estimate this for every galaxy as 
\begin{equation}
\sigma_{\gamma}^2=\sigma_{\gamma,\mathrm{intrinsic}}^2+\sigma_{\gamma,\mathrm{measurement}}^2~,
\end{equation}
where intrinsic shape noise is constant $\sigma_{\gamma,\mathrm{intrinsic}}=0.27$ \citep{2007ApJS..172..219L} and $\sigma_{\gamma,\mathrm{measurement}}$ is derived from the second and fourth order moments weighted by the sum of the variance and absolute value of the sky background \citep{2019arXiv191106333H}. At each step of the iteration, the 2\% most discrepant masses are adjusted. 

\subsubsection{Uncertainty}
\label{sc:uncertainty}
We estimate the noise in each pixel of the mass map via bootstrap re-sampling. To this end, we first select the two $2\,\rm{Mpc}\times2\,\rm{Mpc}$ patches of the sky\footnote{The two patches of sky used to estimate the level of noise in the weak lensing map are centered at ($\alpha=73.644053$, $\delta=-3.012986$) and ($\alpha=73.426295$, $\delta=-3.089766$).} outside the cluster core that contain the smallest grid cells (and hence the highest $K$-band luminosity peaks) where substructures and filaments are most likely to be located. We choose a random orientation for each shear measurement ($\phi$ in equation~\ref{eq:gamma_ini}) in these two patches of sky, then reconstruct a new mass map using \textsc{Lenstool}. Inside an aperture of $r<480$\,kpc, the mean noise level of 100 random realisations is found to be $\langle M\rangle=$ $2.08\times10^{13}\,\msun$, which is non-zero because of \textsc{Lenstool}'s positive-definite mass prior, and its rms uncertainty is $\sigma_M=1.64\times10^{13}\,\msun$. We use the latter to normalise the signal-to-noise ratios of substructures detected in Sect.~\ref{sc:substructures}.

\subsection{The lensing-derived 1D density profile}
\label{sec:1dprofiles}
We calculate the cluster's 1D radial density, $\Sigma(r)$, by (azimuthally) averaging the 2D mass distribution in logarithmically spaced annuli between 80 kpc and 4 Mpc. To enable a statistically rigorous analysis of this key characteristic, we calculate the full covariance matrix $C_{i,j}$ between measurements in each bin (see Sect.~\ref{sec:covariance}).

\subsubsection{Model comparison}

We compare the mean density profile to five models obtained from cosmological simulations: NFW \citep{Navarro:1995iw,Navarro:1996gj}, generalised NFW \citep[gNFW;][]{1996MNRAS.278..488Z}, Einasto \citep{1965TrAlm...5...87E}, Burkert \citep{1995ApJ...447L..25B} and DK14 \citep{Diemer:2014xya}. 
A mathematical definition and description of each halo model is given in Appendix~\ref{sec:appendix:1dprofiles}.
We optimise the free parameters of each model using \textsc{emcee} \citep{2013PASP..125..306F} with a likelihood function 
\begin{equation}
    \mathrm{log}\mathcal{L}=-\frac{1}{2}\sum_{i,j=1}^{N_{\rm{bin}}}(\Sigma_i-\hat{\Sigma}_i)\,C^{-1}_{i,j}(\Sigma_j-\hat{\Sigma}_j)-\frac{1}{2}N_{\rm{bin}}\log{(2\pi |C|)},
    \label{eq:likelihood}
\end{equation}
where $\hat{\Sigma}$ is the model, $N_{\rm{bin}}=20$ is the total number of radial bins, and $|C|$ is the determinant of the covariance matrix. When fitting the NFW, gNFW, Einasto, and Burkert models, we adopt flat uniform priors for $M_{200}\in[5,30]\times10^{14}\,\msun$, and $c_{200}\in[1,10]$. We also adopt a flat prior for the gNFW and Einasto shape parameters, $\alpha\in[0,3]$ and $\alpha_E\in[0.02,0.5]$, respectively. For the Burkert model, we use a flat prior for the core radius, $r_{\rm{core}}\in[100,800]$\,kpc. For the DK14 model, following \citet{More:2016vgs} and \citet{Baxter:2017csy}, we use the priors for $\rho_\mathrm{s}$, $r_\mathrm{s}$, $r_\mathrm{t}$, $\text{log}(\alpha)$, $\text{log}(\beta)$, $\text{log}(\gamma)$, and $s_\mathrm{e}$ that are listed in Table~2 of \citet{Chang:2017hjt}. Because the location of MS\,0451 is so well known from strong-lensing constraints, we omit 
their miscentering term. 

To compare the goodness of fit for models with different numbers of free parameters, we calculate the Bayesian Information Criterion
\begin{equation}
{\mathrm{BIC}}=-2\log{\mathcal{L}} + k\log{N_{\rm{bin}}},
\end{equation}
the Akaike Information Criterion
\begin{equation}
{\mathrm{AIC}} =-2\log{\mathcal{L}} + 2 k,   
\end{equation}
and the corrected Akaike Information Criterion
\begin{equation}
{\mathrm{AICc}} = {\mathrm{AIC}} + \frac{2\,k\,(k + 1)}{(N_{\rm{bin}}-k-1)},   
\end{equation}
where $k$ is the number of free parameters. 
These three information criteria include penalty terms for adding free parameters that make a model more complex. The AIC has a larger penalty term than the BIC; the AICc approaches AIC as $N_{\rm{bin}}$ increases, but is more robust for small $N_{\rm{bin}}$.
For all three criteria, lower values indicate a preferred model.

\subsubsection{Covariance matrix}
\label{sec:covariance}
When fitting a parametric density profile to the azimuthally averaged mass maps, a first estimate of the uncertainty on the density at a given radius can be obtained by looking at the spread of densities at that radius in the MCMC samples generated by Lenstool (or bootstrap sampling, as described in Section~\ref{sc:uncertainty}). However, the non-local mapping between observable shear and reconstructed mass leads to correlations between adjacent pixels. To fully account for these dependencies, we calculate the covariance matrix between radial bins $i$ and $j$
\begin{equation}
    C_{\mathrm{SL+WL}\,(i,j)}=\frac{1}{N}\sum_{l=1}^{N}\big(\Sigma_{l,i}-\langle{\Sigma}_{i}\rangle\big)\big(\Sigma_{l,j}-\langle{\Sigma}_{j}\rangle\big),
    \label{eq:covariance}
\end{equation}
where $N$ is the number of MCMC samples generated by Lenstool, $\Sigma_{l,i}$ is the surface mass density of the $l^\mathrm{th}$ sample in the $i^\mathrm{th}$ spatial bin, and $\langle{\Sigma}_{i}\rangle$ is the mean surface mass density of MCMC samples in the $i^\mathrm{th}$ spatial bin. 

Inside the cluster, the dominant sources of statistical noise are the finite number and unknown intrinsic (unlensed) shapes of the background galaxies used in the weak-lensing analysis. Close to the cluster core, our default joint strong- and weak-lensing analysis underestimates the noise, because we fixed the strong-lensing potentials.
To account for this effect in the covariance matrix, we reconstruct a separate mass map using only weak-lensing information, i.e., we apply the mass grid and shear measurements in the core region too. The resulting covariance matrix $C_{\rm{WL}}$ is valid out to $R\sim1$ Mpc. We then define a combined covariance matrix 

\begin{equation}
   C_{\mathrm{shape}\,(i,j)} =
\begin{cases}
C_{\mathrm{SL+WL}\,(i,j)}+C_{\mathrm{WL}\,(i,j)} & \text{for }R_i~\mathrm{and}~R_j<1\,\text{Mpc} \\
C_{\mathrm{SL+WL}\,(i,j)} & \text{otherwise}.
\end{cases}
\end{equation}
Note that this procedure overestimates $C_{i,j}$ in bins close to the $R\sim 1$ Mpc transition region. However, this effect is negligible in our measurement.

In the outskirts of a cluster, the statistical uncertainty is dominated by large-scale structure (LSS) projected onto the lens plane. While the specific realisation of LSS along the line of sight to MS\,0451 is unknown, we can statistically account for its contribution to the covariance matrix $C_{\mathrm{LSS}\,(i,j)}$ by analysing mock observations of many simulated clusters. In our companion paper \citep{2020arXiv200610155T}, we generate realisations of LSS along 1000 random lines of sight through the BAHAMAS simulation \citep{2017MNRAS.465.2936M}.
We then integrate the 3D mass along the line of sight, weighted by the lensing sensitivity function $\beta(z_l,z_s)=D_{ls}/D_s$ with $\langle z_s\rangle=0.97$, and interpreting it as a mass distribution in a single lens plane at $z_l=0.55$. For each LSS realisation, we calculate an effective radial density profile, $\kappa_{\rm{LSS}}(R)$, with the same radial binning as applied to our data, which allows us to calculate the full covariance matrix, $C_{\mathrm{LSS}\,(i,j)}$, describing LSS at different radii.

Finally, we combine the two components of the covariance matrix across the full range of scales,
\begin{equation}
    C_{(i,j)}=C_{\mathrm{shape}\,(i,j)}+C_{\mathrm{LSS}\,(i,j)}~.
\end{equation}

\subsection{Lensing-derived halo shape}

We measure the shape of the MS\,0451 mass distribution by fitting our reconstructed 2D mass map with elliptical NFW models \citep[eNFW;][]{2010MNRAS.405.2215O}.
We define the centre of the eNFW halo to be the position of BCG ($\alpha=73.545202$, $\delta=-3.014386$), and then optimise\footnote{We use the \textsc{L-BFGS-B} algorithm \citep{53712fe04a3448cfb8598b14afab59b3} from Python's \textsc{scipy.minimize} package \url{https://docs.scipy.org/doc/scipy/reference/generated/scipy.optimize.minimize.html.}} its four free parameters (with the allowed range for the parameters within the optimisation:  
$M_{200}\in[0.5,3]\times 10^{15}\,\msun$, $c_{200}\in[1, 10]$, position angle $\phi\in[0,180]^\circ$, and axis ratio $q=a/b\in[0.1,0.9]$) to minimise the absolute difference between the observed and modelled mass maps, integrated inside a circular aperture. To measure variations in the cluster shape as a function of radius, we repeat this fit inside circular apertures of varying radii. We perform this fit on every mass model created in \textsc{Lenstool}'s MCMC chains, and measure the mean and rms values for each free parameter, marginalising over all others.

\section{METHOD: X-RAY ANALYSIS}
\subsection{X-ray imaging analysis }
\label{sec:method:xray}

We extract \textit{XMM-Newton} images and exposure maps in the [$0.7$--$1.2$]\,keV band from the cleaned event files. To predict the spatial and spectral distribution of the particle-induced background, we analyse a collection of observations performed with a closed filter wheel (CFW). We compute model particle-background images by applying a scaling factor to the CFW data to match the count rates observed in the unexposed corners of the three EPIC cameras. The images, exposure maps and background maps for the three detectors are then summed to maximize the signal-to-noise ratio.

To determine the thermodynamic properties of the cluster, we extract spectra for the three EPIC detectors in seven annular regions from the center of the source to its outskirts (radial range $0\arcmin$--$4\arcmin$). We also extract the spectra of a region located well outside the cluster to estimate the properties of the local X-ray background. The redistribution matrices and effective area files are computed locally to model the telescope transmission and detector response. For each region, the spectra of the three detectors are fitted jointly in {\texttt{XSPEC} \citep{1999ascl.soft10005A}} with a model including the source (described as a single-temperature thin-plasma {APEC model \citep{2001ApJ...556L..91S}} absorbed by the Galactic equivalent neutral hydrogen density $N_H$), the local three-component X-ray background as fitted in the background region, and a phenomenological model tuned to reproduce the spectral shape and intensity of the particle background. The best-fitting parameters of the APEC model (temperature, emission measure, and metal abundance) as a function of radius are obtained by minimizing the $C$ statistic. 

\subsection{X-ray 1D surface-brightness profile}
\label{sec:xray1d}

To measure the 1D surface-brightness profile of the cluster, we use the azimuthal median technique \citep{Eckert2015}, which allows us to excise contributions from infalling substructures and asymmetries. To this end, we employ Voronoi tessellation to create an adaptively binned surface-brightness map of the cluster. For each annulus, we then determine the median value of the distribution of surface-brightness values. Uncertainties are estimated by performing $10^4$ bootstrap resamplings of the distribution and computing the root-mean-square deviation of the measured medians. We measure the local background outside the cluster, where the brightness profile is flat, and subtract it from the source profile. Finally, gas-density profiles are determined by deprojecting the surface-brightness profile, assuming spherical symmetry. 

We estimate the mass profile of the cluster from the gas-density and temperature profiles, assuming HSE \citep[see][for a review]{pratt19}, i.e., that the pressure gradient balances the gravitational force:
\begin{equation}
    \frac{dP_{\rm gas}}{dr} = -\rho_{\rm gas}\frac{GM(<r)}{r^2}.
    \label{eq:hse}
\end{equation}
The profile of the gravitational mass can thus be inferred from the gas-pressure and density profiles. To solve equation~\eqref{eq:hse}, we use the \emph{backwards} approach introduced by \citet{Ettori2019} which adopts a parametric model for the mass profile (here, NFW) and combines it with the density profile computed through the multiscale decomposition technique to predict the pressure (and hence, temperature) as a function of radius. The model temperature profile is projected along the line of sight and corrected for multi-temperature structure using scaling as described in \citet{Mazzotta04}. The projected temperature profile is then compared to the data, and the parameters of the mass model (i.e., mass and concentration) are optimized using MCMC. The integration constant, 
which describes the overall pressure level at the edge, is left free while fitting and determined on the fly. For more details on the mass reconstruction technique and a careful assessment of systematic effects and uncertainties, see \citet{Ettori2019}. 

The cumulative gas-mass profile is computed by integrating the gas-density profile over the cluster volume, assuming spherical symmetry:
\begin{equation}
    M_{\rm gas}(<r) = \int_0^r 4\pi r^{\prime\ 2} \rho_{\rm gas}(r^{\prime})\,dr^{\prime}.
\end{equation}
\noindent Here $\rho_{\rm gas}=\mu m_p (n_e+n_H)$, with $n_e$ and $n_H=n_e/1.17$ being the number density of electrons and protons, respectively; $\mu=0.61$ is the mean molecular weight, and $m_p$ is the proton mass. Our procedure directly computes the hydrostatic gas fraction $f_{\rm gas,HSE}(r)=M_{\rm gas}(r)/M_{\rm HSE}(r)$, which traces the virialization state of the gas \citep{Eckert19}.

\begin{figure*}
\centering
\includegraphics[width=1\textwidth,trim={0 2mm 0 2mm},clip]{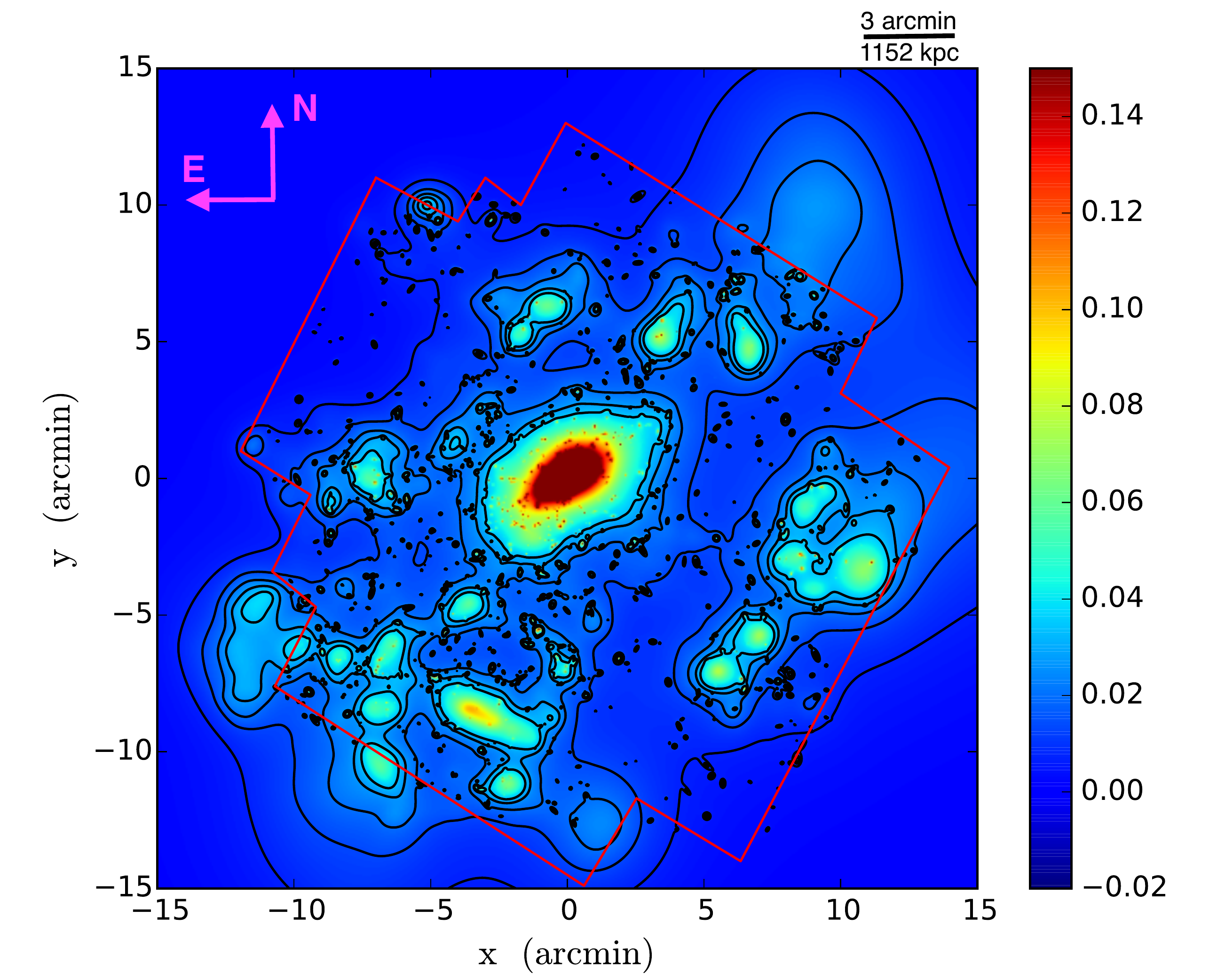}
\caption{
The projected distribution of mass around MS\,0451, inferred from our \textsc{Lenstool} strong- and weak-lensing reconstruction and centred on the BCG ($\alpha$$=$$73.545202$, $\delta$$=$$-3.0143863$). 
{Colours indicate the projected convergence, $\kappa$}.
Black contours mark the signal-to-noise ratio in steps of $1\sigma_\Sigma$, measured from bootstrap re-sampling (see Sect.~\ref{sc:uncertainty}). The red polygon delineates the field of view of the \emph{HST}/ACS imaging mosaic. } 
\label{fig:ms0451_massrec}
\end{figure*}

\section{Results} 


The mass distribution around MS\,0451 (shown in Fig.~\ref{fig:ms0451_massrec}) has a core that is elongated along an axis from South-East to North-West and surrounded by lower-mass substructures.
An alternative reconstruction using  direct inversion following \citet{1993ApJ...404..441K} finds consistent features (Appendix~\ref{sec:massrec_mrlens}).
However, our primary \textsc{Lenstool} method achieves higher spatial resolution in regions containing cluster member galaxies and is more efficient at suppressing noise in regions without them.  

\begin{figure}
\centering
\includegraphics[width=0.5\textwidth,trim={0 2mm 0 3mm},clip]{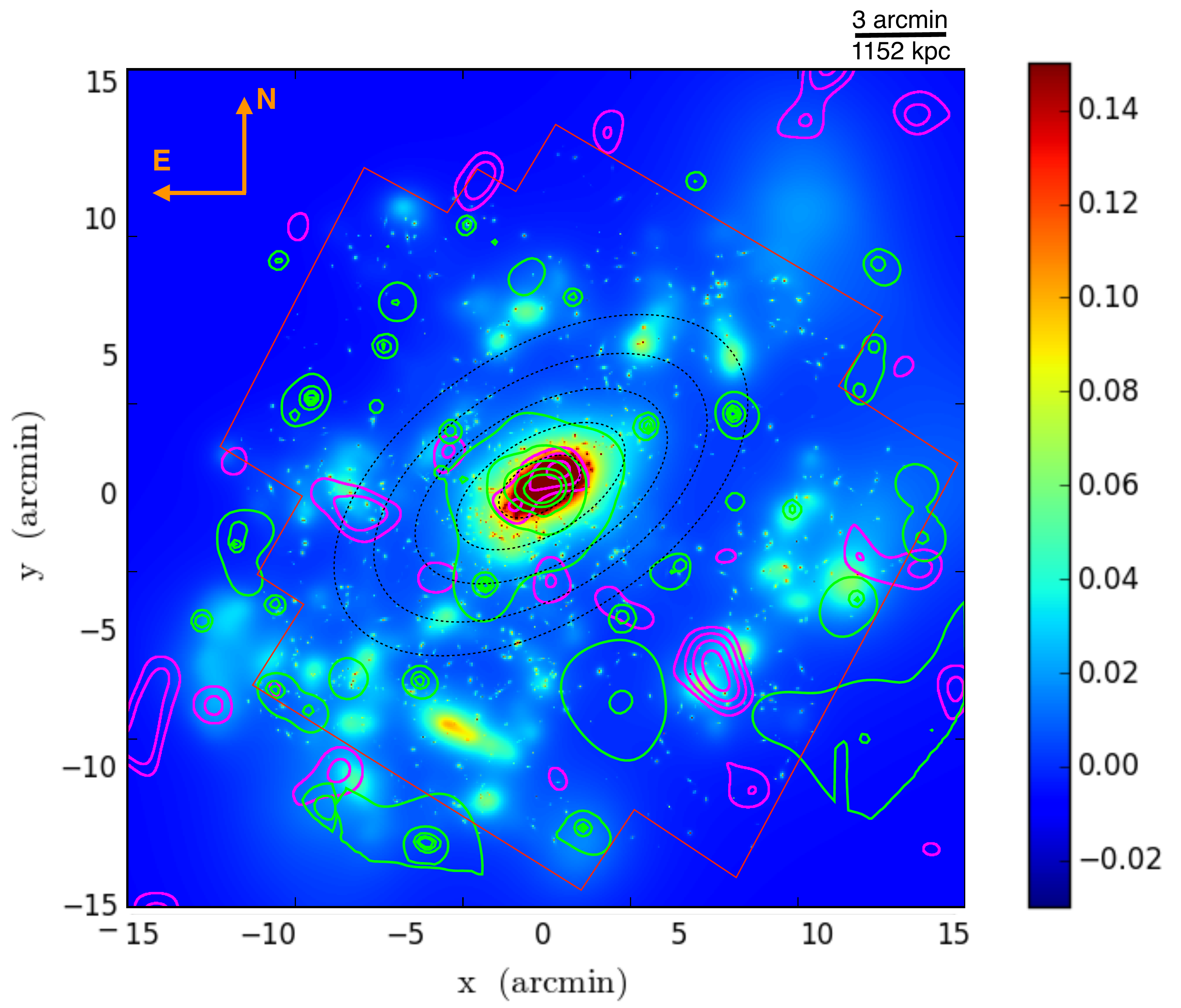}
\caption{
Alternative probes of the mass distribution around MS\,0451, overlaid for ease of reference on the colour image from Fig.~\ref{fig:ms0451_massrec}. 
Magenta contours show weak-lensing measurements from ground-based observations (private communication N.\ Martinet), starting at $3\sigma_{\kappa}$ and in steps of $1\sigma_{\kappa}$, the rms uncertainty on convergence. 
Green contours show the X-ray surface brightness as recorded by {\it XMM-Newton}. 
Black ellipses show the shape of the eNFW model that best fits our {\sc Lenstool} reconstruction within circular apertures of different radii (defined by the semi-major axis).}
\label{fig:xray_contour}
\end{figure}

The elongated core, which contains two distinct mass peaks, is consistent with an analysis of CFHT/Megacam ground-based weak-lensing measurements \citep[][and shown in Fig.~\ref{fig:xray_contour} with magenta contours provided via private communication by N.\ Martinet]{2016A&A...590A..69M}.
We confirm the existence of several nearby substructures, but our higher S/N data do not show them connected into a filament running South-West to North-East, as hypothesised by \cite{2016A&A...590A..69M}.
X-ray emission is detected out to $R=1.7$\,Mpc (Fig.~\ref{fig:xray_contour}).

\subsection{Total mass and density profile}
\label{Sec:densityprofile}

\begin{figure}
\centering
\includegraphics[width=0.53\textwidth,trim={0 2mm 0 10mm},clip]{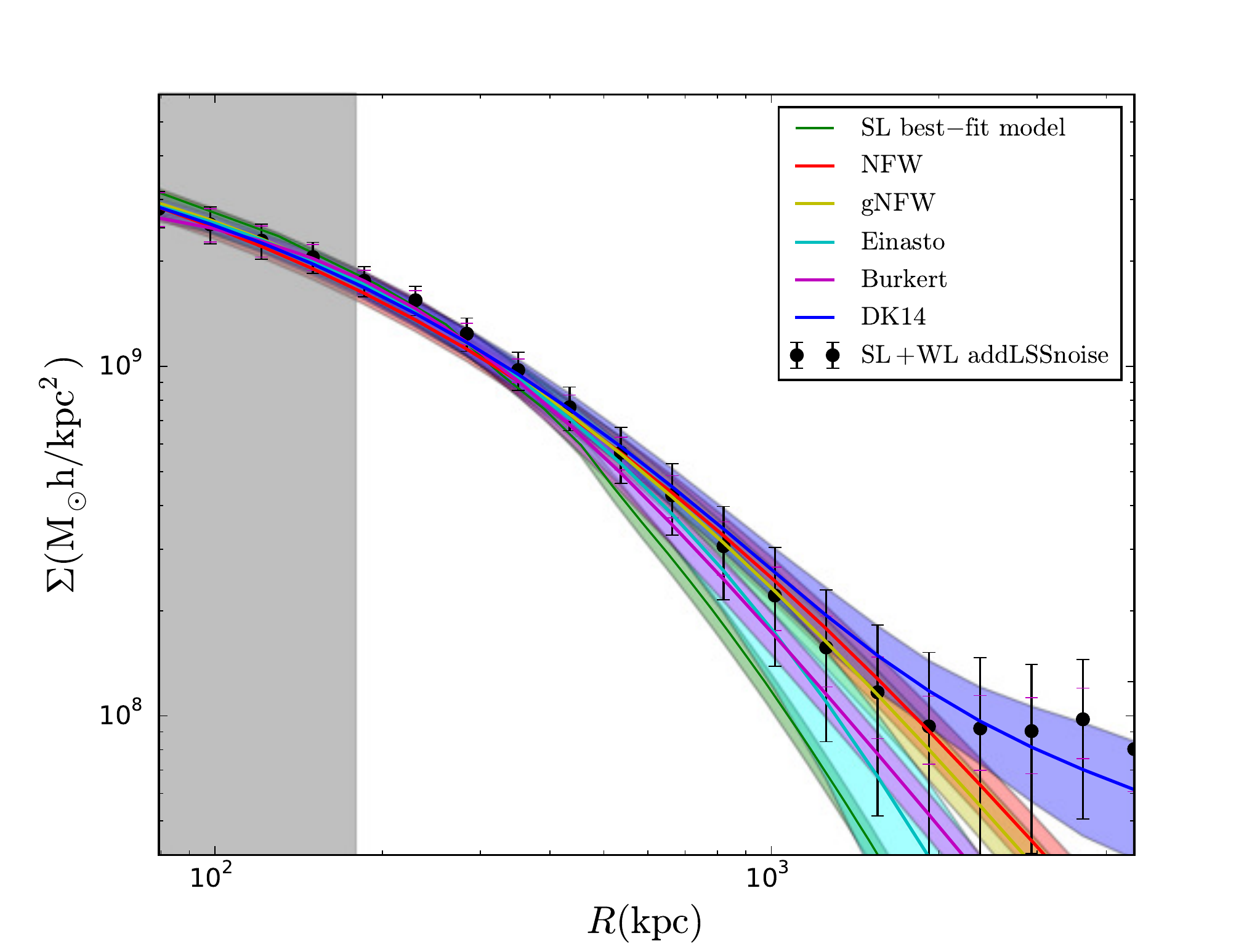}
\caption{Azimuthally averaged 1D profile of mass in MS\,0451 (black data points), from our combined strong- and weak-lensing analysis (Fig.~\ref{fig:ms0451_massrec}). The double error bars show the statistical uncertainty caused solely by the galaxies' intrinsic shapes (inner error bar) and the uncertainties when line-of-sight substructures are also taken into account (outer).
The green curve shows the best-fit model using only strong-lensing information \citep{2020arXiv200610700J}, extrapolated beyond the multiple-image region (grey shaded area). 
Solid lines in other colours and their respective shaded areas show the mean and 68\% confidence intervals from fits to various models.}
\label{fig:densityprof}
\end{figure}

Our combined weak- and strong-lensing reconstruction smoothly extends the surface mass density profile outside the multiple-image region (Fig. ~\ref{fig:densityprof}).
We measure a projected mass $M(R<195\,\text{kpc})=(1.85\pm0.87)\times10^{14}\,\msun$, consistent with previous strong-lensing measurements of $1.73\times10^{14}\,\msun$ \citep{2010A&A...509A..54B} and $1.8\times10^{14}\,\msun$ \citep{2011MNRAS.410.1939Z}.
At larger radii, our analysis is sensitive for the first time to additional infalling or projected substructures; compared to previous models, based solely on strong-lensing features, we detect excess mass at $R>3$\,Mpc. 

Theoretically motivated models to fit the 1D lensing signal (Fig. \ref{fig:densityprof}) are described in Appendix~\ref{sec:appendix:1dprofiles}, and their best-fit parameters are listed in Table~\ref{tab:parameters}.
For the best-fit NFW model we measure a mass  $M_{200c}=(1.65\pm0.24)\times10^{15}\,\msun$ inside $R_{200c}=1.99\pm0.06\,$Mpc, or $M_{500}=(1.13\pm 0.16)\times10^{15}\,\msun$, and concentration $c_{200}=3.79\pm0.36$. Within the statistical uncertainty, this result is consistent with the ground-based weak-lensing measurement of $M_{200}=(1.44^{+0.33}_{-0.26})\times10^{15}\,\msun$ for fixed $c_{200}=4$ \citep{2012A&A...546A.106F}. 
{The Burkert profile is disfavoured by the BIC and AIC. 
For the best-fit gNFW and Einasto models we find masses and concentrations slightly lower than for NFW. However, their BIC and AIC differ by less than $2$ from the NFW ones. Thus, we conclude that our data are unable to distinguish between these three models with statistical significance (as quantified by \citealt{ModelSelection}).} We therefore adopt the NFW model as our default in the following analysis.

We note that the additional complexity of the DK14 model captures a splashback-like feature at $R\sim$2\,Mpc (see Appendix~\ref{sec:appendix:splashback}). However, the BIC and AIC both disfavour the DK14 model, and the mentioned  feature might be caused by noise or the projection of unrelated large-scale structure along the line of sight.

\begin{table*}
    \centering
    \caption{Marginalized posterior constraints on cluster model parameters, and the differences between their information criteria and those of the best-fit NFW model. The information criteria of an NFW  are BIC$_{\rm{NFW}}$=13934.50, AIC$_{\rm{NFW}}$=13932.51 and AICc$_{\rm{NFW}}$=13933.22. Lower values indicate preferred models.
    }
    \fontsize{11pt}{12pt}\selectfont
    \begin{threeparttable}[t]
    \begin{tabular}{lcccccc}
    \hline
    \hline
    {Models} & {$M_{200c}[10^{14}\,\msun]$} & {$c_{200}$} & {Shape parameter}& {$\Delta$BIC}& {$\Delta$AIC}& {$\Delta$AICc}\\
    \midrule
    NFW &   $16.51\pm2.44$ & $3.79\pm0.36$ & -- &0 &0 &0 \\
    gNFW &  $16.10\pm1.94$ & $4.47\pm0.47$ & $\alpha=0.57\pm0.20$& 0.51  & -0.48 & 0.02 \\
    Einasto &  $14.32\pm2.67$ & $4.26\pm0.50$ & $\alpha_E=0.42\pm0.11$ & 0.02  &-1.00 & -0.51\\
   {Burkert} & {$13.62\pm1.62$} & & {$r_\mathrm{core}=(230\pm20)$\,kpc} & {7.60} & {6.60} &  {7.10} \\
    DK14\tnote{a} &  &  &  &9.60  & 4.60  & 12.90 \\
    \midrule
    X-ray & $17.47\pm7.61$ & $2.35_{-0.70}^{+0.89}$ & -- \\
    \hline
    \hline
    \end{tabular}
    \begin{tablenotes}\small
     \item[a] Parameters of the DK14 model are excluded from this table for clarity. These are listed in Table~\ref{tab:DK14tab}.
   \end{tablenotes}
    \end{threeparttable}
    \label{tab:parameters}
\end{table*}

\subsection{Halo shape}
\label{Sect:haloshape}

We measure the projected shape of MS\,0451 by fitting the 2D mass distribution inside a circular aperture with an eNFW model. This approach yields results that are consistent with the previous 1D fit (Sect.~\ref{Sec:densityprofile}): for the region inside $R<3.24$\,Mpc, we obtain $M_{200c}=(1.57\pm0.14)\times10^{15}\,\msun$ and $c_{200}=3.7\pm0.4$. 
The best-fit {axis ratio $q=b/a$} varies as a function of radius, from $q=0.48\pm0.01$ within $R<649$\,kpc to $q=0.57\pm0.03$ within $R<3.24$\,Mpc. The cited statistical uncertainty may be an underestimate because we have neglected correlations between adjacent pixels in our error model of the mass map and, in the cluster core, because of our use of fixed strong-lensing potentials during MCMC parameter search. The axis ratio is consistent with simulations of general clusters \citep[][]{2020arXiv200610155T,Jing:2002np,Suto:2016zqb}, but smaller than the value of $q=0.72$ ($649$\,kpc\,$<$$\,R\,$$<974$\,kpc) measured from ground-based lensing observations by \cite{2015A&A...581A..31S}. This discrepancy might be explained by the large smoothing kernel used by Soucail and coworkers to reconstruct the mass distribution, which artificially circularises the data. 
Indeed, our results more closely resemble those from lensing analyses of large cluster samples, including \cite{2010MNRAS.405.2215O}, who found $\langle q\rangle=0.54\pm0.04$ for 18 X-ray luminous clusters at $0.15<z<0.3$, and \cite{Umetsu:2018ypz}, who found $\langle q\rangle=0.67\pm0.07$ for the CLASH sample of 20 massive clusters. 

At all radii, we find that MS\,0451 is elongated roughly along a North-West to South-East axis (Fig~\ref{fig:xray_contour}), with a mean orientation $\sim$$31.9^\circ$ counter-clockwise from East. The $\sim$$10\%$ variation in this angle between the inner ($R<640$\,kpc) and outer halo ($R<3.24$\,Mpc) agrees well with typical clusters in both simulations \citep{Despali:2016pkj} and observations \citep{2019arXiv191106333H}.

\subsection{Baryonic components}

\subsubsection{Distribution of baryons}
\label{sec:ICM}

\begin{figure}
\centering
\includegraphics[width=0.48\textwidth]{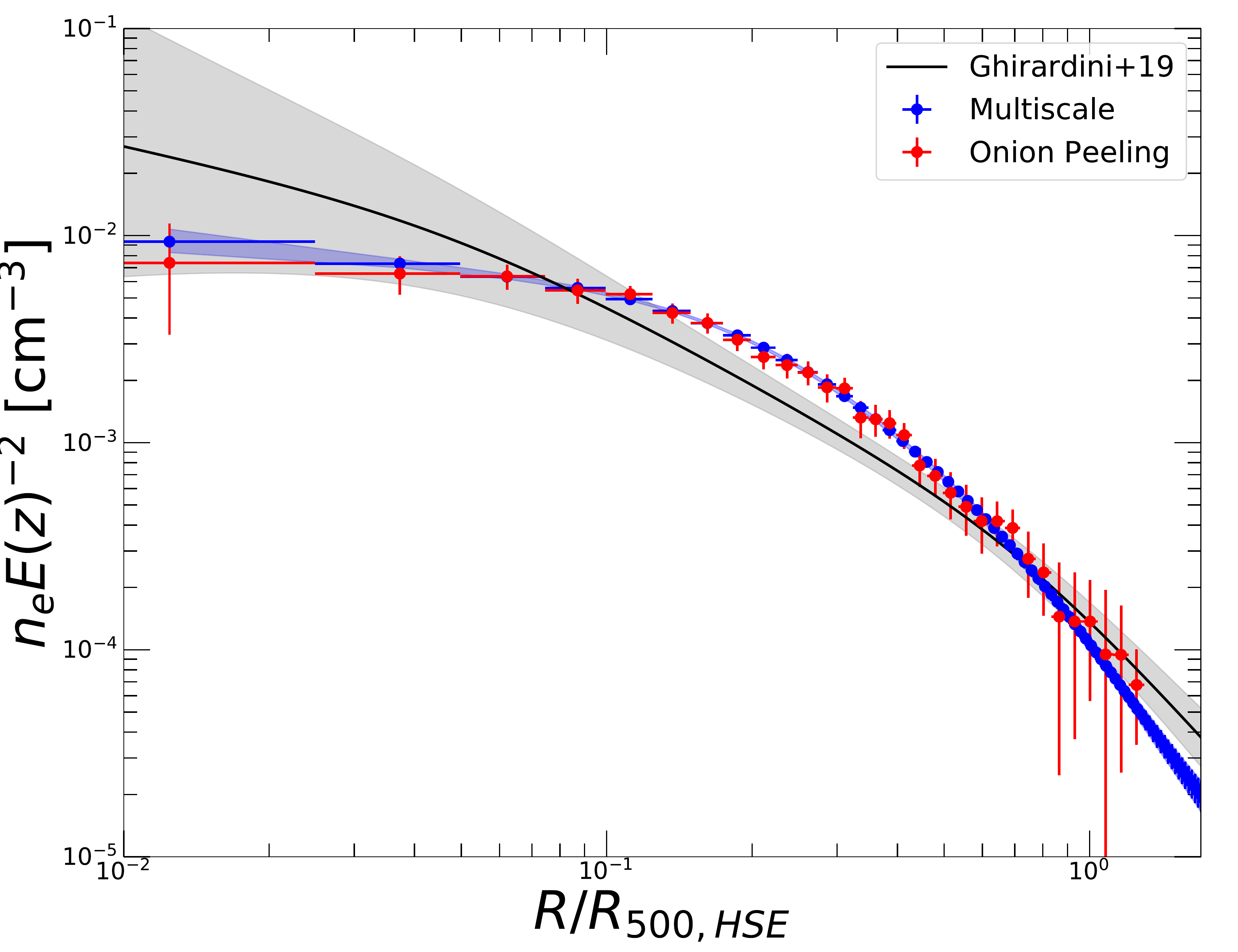}
\includegraphics[width=0.48\textwidth]{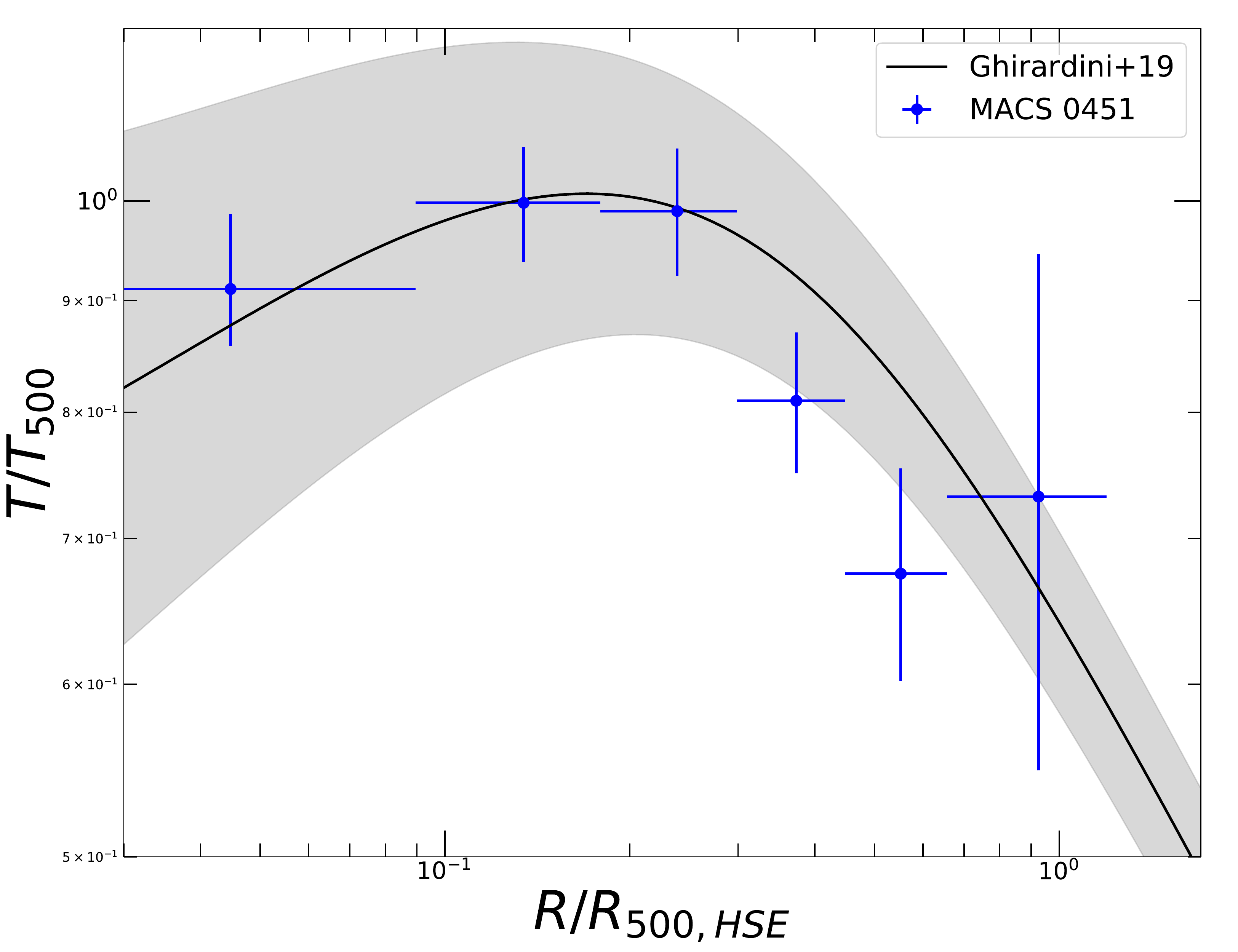}
\caption{Thermodynamic profiles of MS\,0451's Intra-Cluster Medium (ICM), scaled according to the self-similar model \citep{Kaiser1986}. \emph{Top:} Deprojected electron-density profile of the cluster computed using the onion peeling (red), and multiscale decomposition (blue) methods. 
\emph{Bottom:} Spectroscopic-temperature profile of the cluster (blue). In both panels, the black curve and gray shaded areas show the mean profile and $1\sigma$ scatter of the X-COP sample of massive clusters at low redshift \citep{Ghirardini2019} for comparison. 
} 
\label{fig:thermo_profiles}
\end{figure}

To measure the cluster's electron-density profile, we apply the non-parametric ``onion peeling'' algorithm \citep{Kriss1983} and the multiscale decomposition technique \citep{Eckert2016} to the X-ray data (Fig.\ \ref{fig:thermo_profiles}). Both methods assume spherical symmetry, and both yield consistent results. We find the distribution of baryons to be different from the one found by our lensing analysis, in that it shows a constant-density core, flatter than both our lensing results and the distribution of gas in a typical massive cluster from the X-COP low-redshift sample \citep{Ghirardini2019}.

To measure the cluster's temperature profile, we fit a single-temperature plasma-emission model to the X-ray spectra extracted in six concentric annuli spanning the radial range 0--1.5\,Mpc. 
We find that the temperature of the X-ray emitting gas decreases from $\sim$9\,keV in the core to $\sim$6\,keV in the outskirts, consistent with the `universal' thermodynamic profile of X-COP clusters.

In a separate analysis of the X-ray data assuming hydrostatic equilibrium and spherical symmetry, we measure a hydrostatic mass of $M_{500,\rm{HSE}}=(1.06\pm0.35)\times10^{15}\,\msun$, and a concentration $c_{200,\rm{HSE}}=2.35^{+0.89}_{-0.70}$. The concentration is again lower than the one we obtain in our lensing analysis. Extra\-polating the best-fit model to large radii yields a total mass of $M_{200c,\rm{HSE}}=(1.75\pm 0.75)\times10^{15}\,\msun$, which inevitably has large uncertainties because the X-ray emission at these radii is faint. 

We note that the assumption of hydrostatic equilibrium may not be appropriate for this cluster. Deeper X-ray imaging and/or constraints on the Sunyaev-Zel'dovich signal are required to quantify the level of non-thermal pressure support.

\begin{figure}
    \centering
    \includegraphics[width=0.48\textwidth]{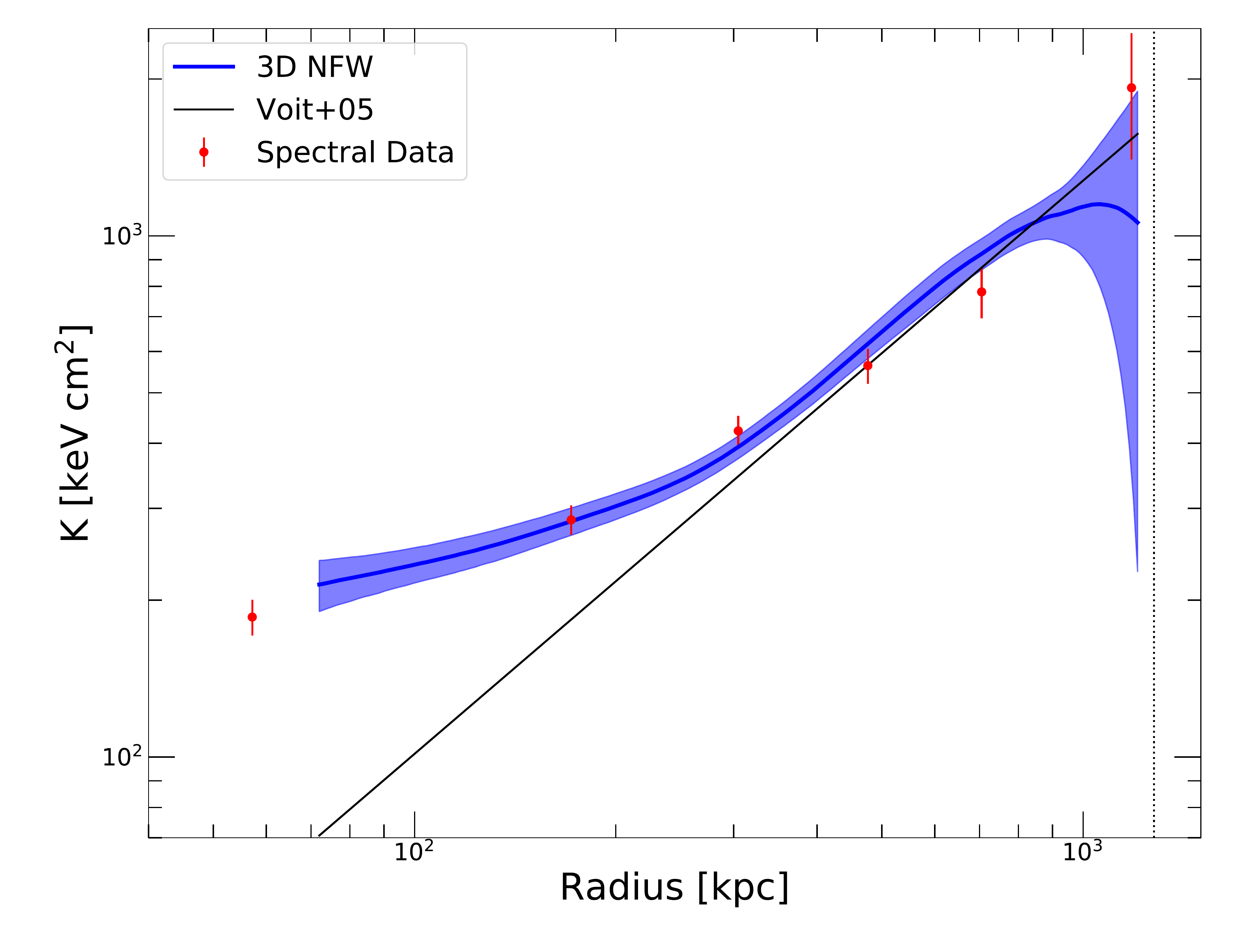}
    \caption{Radial profile of the gas entropy. The red data points are obtained from the measured spectroscopic temperature and the gas density. The blue curve is the model optimised using the \emph{backwards fit} method. For comparison, the black curve shows the gas entropy predicted by the \citet{2005RvMP...77..207V} gravitational-collapse model.}
    \label{fig:gas_entropy}
\end{figure}

The radial entropy profile of the intra-cluster gas (Fig.~\ref{fig:gas_entropy}), obtained by combining the measured spectroscopic temperature with the gas density, is consistent with the 3D entropy model recovered from the \emph{backwards NFW fit} under the assumption of HSE. We find a strong entropy excess in the cluster core, compared with the entropy of the fully relaxed gas calculated from the gravitational-collapse model \citep{2005RvMP...77..207V}. This large entropy excess confirms that MS\,0451 does not contain a cool core.

\subsubsection{Baryonic-mass fraction}

To measure the gas-mass fraction $f_{\mathrm{gas}}$, we first integrate the non-parametric gas profiles (which do not assume hydrostatic equilibrium) and obtain a total gas mass of $M_{{\mathrm{gas}},500}=(1.29\pm 0.15)\times10^{14}\,\msun$ inside a sphere of radius $R_{500,\rm{HSE}}=1.28\pm0.14$\,Mpc. Division by the total mass $M_{500}$ of the NFW model, which best fits the lensing data inside a sphere of radius $R_{500}=(1.30\pm0.06)$ Mpc, yields  $f_{\mathrm{gas},500}=(11.6\pm2.1)$\%, in good agreement with the result of $f_{\rm gas,500,HSE}=(12.2\pm4.3)\%$ from our analysis assuming hydrostatic equilibrium.

To measure the stellar-mass fraction, we use the ratio of stellar mass to light of quiescent galaxies 
\begin{equation}
\text{log}_{10}\left(M_{*}/L_K\right)=a\,z+b,
\end{equation}
where $a = -0.18 \pm 0.04$ and $ b = +0.07 \pm 0.04$ 
\citep{2007A&A...476..137A}, assuming a \cite{1955ApJ...121..161S} initial mass function (IMF)\footnote{To convert from Salpeter to a \cite{2003PASP..115..763C} IMF, we adjust the stellar masses by 0.25\,dex and find $f_{\ast,500}=(1.6\pm0.24)\%$, similar to $f_{\ast}$$\sim$$1.5\%$ measured in the wide-field \emph{HST} COSMOS survey \citep{2012ApJ...744..159L}.}. Applying this relation to all 1277 cluster member galaxies in the \emph{HST}/ACS mosaic yields a mean value of $\langle M_{*}/L_K\rangle=0.94\pm0.003$ and a total stellar mass of  ${M_{*,500}}=(3.37\pm0.03)\times10^{13}M_\odot$, where the uncertainty is the error of the mean. Note that, although the integration is performed over a cylinder of radius $R_{500}$ rather than a sphere, the result is largely unaffected since the stellar mass is extremely centrally concentrated. We thus adopt a stellar-mass fraction of $f_{\ast,500}=(3.0\pm0.4)\%$. 

\label{sec:fbar}

Combining these measurements, we obtain a total baryonic-mass fraction, $f_{\rm b,500} \equiv f_{\ast,500} + f_{\rm gas,500} = (14.6\pm1.4)\%$. This value is consistent with the mean cosmic baryon fraction of $f_{\rm b}=(14\pm2)\%$ measured from the outskirts of clusters at $z<0.16$ \citep{Mantz:2014xba}, and also with $f_{\rm b}=(15.6\pm0.3)\%$ from the Cosmic Microwave Background \citep{2016A&A...594A..13P}. 

\begin{table*}
  
  \label{tab:table1}
  \resizebox{\textwidth}{!}{%
  \begin{tabular}{ccccS[table-align-uncertainty,table-figures-uncertainty=1]S[table-align-uncertainty,table-figures-uncertainty=1]c}
   \hline
    \hline
    {Substructure} & {R.A.} & {Dec.} & ${\langle z\rangle}$& {$M_{\rm{tot}}$[$10^{13}\,\msun$]}& {$M_{\rm{stellar}}$ [$10^{11}\,\msun$]} & {Detection S/N}\\
    \midrule
    Sub1 & 4:54:26.917 & -2:59:39.894 & 0.62 &  6.17\pm 2.70 &  6.59\pm 1.79 & 3.76 \\
    Sub2 & 4:54:39.389 & -3:00:32.808 & 0.58 &  8.34\pm 3.30 & 12.70\pm 2.26 & 5.09\\
    Sub3 & 4:54:15.278 & -3:03:11.620 & 0.61 & 13.50\pm 2.90 & 31.25\pm 3.04 & 8.23\\
    Sub4 & 4:54:26.088 & -3:05:37.949 & 0.63 &  7.17\pm 3.23 &  5.83\pm 1.56 & 4.37 \\
    Sub5 & 4:54:11.745 & -3:07:30.042 & 0.55 &  6.12\pm 2.96 & 10.55\pm 2.13 & 3.73 \\
    Sub6 & 4:54:37.972 & -3:07:33.134 & 0.56 &  8.42\pm 3.37 &  8.05\pm 2.28 & 5.13\\
   \hline
   \hline
  \end{tabular}}
  \caption{Confirmed substructure detections in MS\,0451 containing cluster member galaxies with a detection S/N of at least 3. Columns show the location of each mass peak, the mean redshift of member galaxies within a 480\,kpc aperture, the lensing and stellar masses integrated within the same aperture, and the signal-to-noise ratio of detection, using the mean noise level of the mass map (Sect~\ref{sc:uncertainty}).
  \vspace{-3mm}}
  \label{tab:substr}
\end{table*}

\subsection{Group-scale substructures}
\label{sc:substructures}

\begin{figure}
\centering
\includegraphics[width=0.5\textwidth,trim={0 1mm 0 1mm},clip]{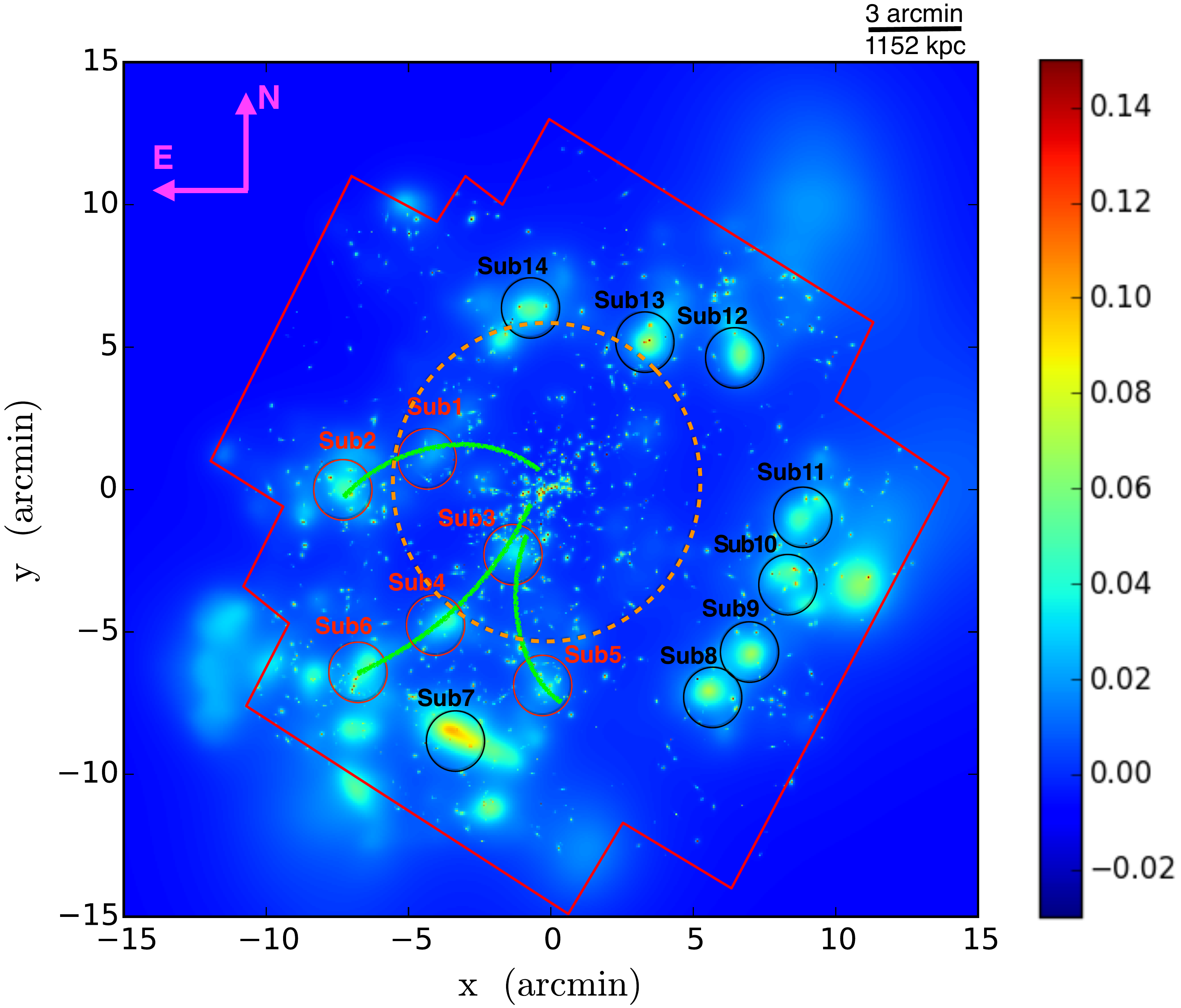}
\caption{The low density environment surrounding MS\,0451.
The colour image shows lensing convergence with SL potentials subtracted: all the remaining signal was constrained by the potential grid and cluster member galaxies. 
The dashed orange circle has radius $R_{\rm{200c}}=1.99\,$Mpc. 
Smaller circles (with radius $480$\,kpc) mark substructures with a projected mass $>3\Sigma_M$ inside that aperture; red circles have optical counterparts at the cluster redshift. 
Green lines suggest the extent and direction of possible large-scale filaments.}
\label{fig:MS0451_subs}
\end{figure}

To study the low-density environment of large-scale structures surrounding MS\,0451, we subtract the strong-lensing potentials from the {\sc Lenstool} convergence map (Fig~\ref{fig:MS0451_subs}). Outside the main halo, {we detect 14 weak-lensing peaks with S/N$>$3 integrated within circular apertures of radius $R=480$\,kpc.} To determine whether these 14 overdensities are at the redshift of the cluster, we assess the redshift distribution of all galaxies within those apertures that have spectroscopic or (mainly) photometric redshifts (Appendix~D). The redshift distribution of galaxies along the line of sight to Substructures 1, 2, 3, 4, 5 and 6 peaks in the range $0.48<z<0.61$. We thus infer that these structures are part of the extended cluster, while all others are projections of structures at other redshifts along our line of sight. The total masses and stellar masses of the six substructures likely to be associated with MS\,0451 are listed in Table~\ref{tab:substr}.

Previous ground-based weak-lensing analyses identified only Substructures 1 and 2 \citep{2016A&A...590A..69M}, or Substructure 2 at a modest 2$\sigma$ significance \citep{2015A&A...581A..31S}. Our identification of 12 significant new structures demonstrates the unique ability of space-based imaging to detect weak-lensing signals in low-density environments.

The mismatch between the structures uncovered by our lensing and X-ray analyses is puzzling. Although the depth of the existing \emph{XMM-Newton} data should be sufficient to detect $\sim10^{14}\,\msun$ halos, we detect faint X-ray emission only from Substructure 6. Brighter -- but misaligned -- X-ray emission is seen near Substructures 2 and 5, and between Substructures 3 and 4. The discrepancy might be caused by selection biases in our analyses. On the lensing side, the high mass of Substructure 3 ($M_{\rm{tot}}=(1.3\pm0.3)\times10^{14}\,\msun$ inside a 480\,kpc aperture) might be erroneous and caused by the structure's proximity to the cluster core. If the main cluster is imperfectly modelled and subtracted, its residual projected mass could artificially boost the lensing signal of Substructure 3. Indeed, all substructures are closer to the cluster's major axis than to its minor axis, and hence the lensing signal from all of them could be biased high. Conversely, proximity to the cluster core also results in a high X-ray background, which lowers the signal-to-noise ratio of the X-ray emission and thus raises the detection threshold. An alternative, physical explanation for the low X-ray emission from these substructures could be that, within $R_{\rm{200m}}=(2.51\pm0.14)$ Mpc, they probably also lie within the 3D splashback radius of MS\,0451 \citep{2015ApJ...810...36M} and thus may have already passed through pericentre. Ram-pressure stripping during the passage through the main halo could have removed much of their hot gas and thus reduced their X-ray luminosity.

\subsection{Filaments}
\label{sc:filament}

\subsubsection{Alignments of substructures}

Based on the distribution of substructures around MS\,0451, we propose that three filaments are connected to the cluster core (shown as green lines in Fig.~\ref{fig:MS0451_subs}). The first of these possible filaments extends East of the cluster, encompassing Substructures 1 and 2 and containing mean convergence $\langle\kappa\rangle=0.022\pm0.006$.
The second points South-East, encompassing Substructures 3, 4, and 6, with $\langle\kappa\rangle=0.033\pm0.007$. The third, finally, turns South, from Substructure 3 to Substructure 5 and also has mean convergence  $\langle\kappa\rangle=0.033\pm0.007$.
For each of these three candidate filaments the density contrast exceeds the threshold value of $\kappa=0.005$ defined in our companion paper \citep{2020arXiv200610155T} to identify filaments, and each has a mean {\it excess convergence} greater than $0.02$, even after subtracting the smooth, cluster-scale mass distribution.

All three possible filaments point in a similar direction, close to the main cluster's South-East/North-West major axis. We detect no substructures in the opposite direction along the same axis (with the possible exception of an unconvincing feature just outside the \emph{HST} mosaic to the North-West). This is strikingly different from the typical distribution of mass in cosmological simulations, which usually show a symmetry of infalling material along both directions of a cluster's major axis, as the system grows and becomes increasingly elongated as the result of gradual, continuous accretion along filaments.

\subsubsection{Aperture multipole moments}

\begin{figure}
\includegraphics[width=0.5\textwidth]{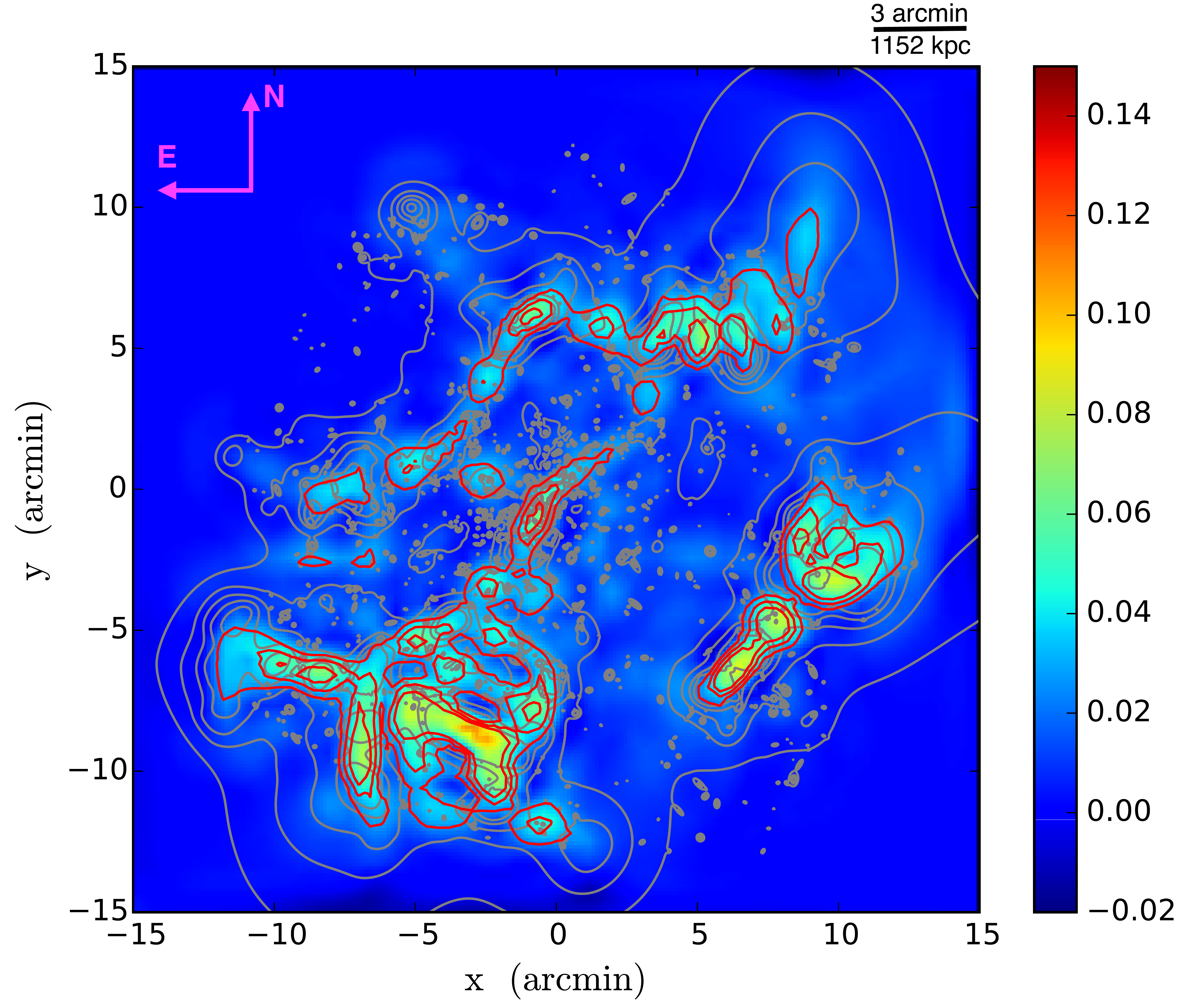}
\caption{The low-density environment around MS\,0451 (Fig.~\ref{fig:MS0451_subs}), filtered using aperture multipole moments to search for extended, filamentary structures. 
Red contours indicate the signal-to-noise ratio, starting at 2 and increasing in steps of 1.
Grey contours correspond to the unfiltered mass distribution.
}
\label{fig:Q_MS0451}
\end{figure}

Extended structures can also be identified through measurements of aperture multipole moments (AMMs) of the 2D mass distribution \citep{Schneider:1996yy}, defined as
\begin{equation}
    Q^{(n)}(\boldsymbol{R})=\int_0^{\infty}\int_0^{2\pi} |\boldsymbol{R}'-\boldsymbol{R}|^{n+1}\, \mathrm{e}^{ni\phi}~U(|\boldsymbol{R}'-\boldsymbol{R}|)~\kappa(\boldsymbol{R}')~\mathrm{d}\boldsymbol{R}'\mathrm{d}\phi\,,
    \label{eq:multipoleint}
\end{equation}
where $n$ is the order of the multipole, ($R$, $\phi$) are polar coordinates, and $U(R)$ is a radially symmetric weight function with characteristic scale $R^{(n)}_{\mathrm{max}}$.
In tests using mock observations of ten massive simulated clusters with $M_{200}\sim10^{15}\msun$ at $z=0.55$ \citep{2020arXiv200610155T}, 
we developed a combination of AMMs that highlights the signal from extended filaments,
\begin{equation}
Q\equiv\alpha_0Q^{\rm{(0)}}+\alpha_1 Q^{\rm{(1)}}+\alpha_2 Q^{\rm{(2)}},
\end{equation}
with optimised constants $\alpha_0$$=$$-\alpha_1$$=$$0.7$, $\alpha_2$$=$$1$, and $R^{(0)}_{\mathrm{max}}$$=$$1\arcmin$, $R^{(1)}_{\mathrm{max}}$$=$$R^{(2)}_{\mathrm{max}}$$=$$2\arcmin$.
This choice of constants enables the detection of narrow filaments with a purity of greater than 75\% and completeness in excess of 40\%. The quadrupole term, $Q^{\rm{(2)}}$, is sensitive to linearly extended mass distributions. The dipole term $Q^{\rm{(1)}}$ fills in the rings which are added around isolated substructures, and the monopole term $Q^{\rm{(0)}}$ suppresses signal between structures. To avoid the massive cluster halos that would dominate the contrast in the low-density filaments, we apply this filter to the convergence map of MS\,0451 after subtracting the strong-lensing component. The resulting $Q$ map is shown in Figure~\ref{fig:Q_MS0451}. We quantify the level of noise by defining $\sigma_Q$ as the standard deviation of all pixels in the $Q$ map.

Although the signal-to-noise ratios for $Q$ are low, the three possible filaments proposed in Sect.~\ref{sc:filament} are also highlighted by the AMM filter, with signal-to-noise ratios of $\sim2-3$. Additional extended structures may exist at other redshifts. In particular, Substructures 8, 9, and 11 might form a linked system at $z\sim0.7$ behind the cluster (Appendix~D). 

While the results presented in Figure~\ref{fig:Q_MS0451} are promising, the limiting, single-orbit depth of the available \emph{HST} observations prevents us from draw firm conclusions from this measurement. Deeper space-based observations obtained by future surveys will increase the density of detected background galaxies and reduce the noise level of this technique.

\section{Discussion: inferred dynamical state}
\label{sec:dynamical}

According to $N$-body and hydrodynamical simulations \citep[e.g.][]{Ritchie:2001qt,Poole:2006nm,Nelson2011}, mergers between two clusters (classified by their masses as the primary and secondary component) proceed in five distinct stages: pre-interaction, first core-core interaction, apocentric passage, secondary core accretion, and relaxation. After the first core-core interaction, the gas of the two merging halos (including the remaining part of the secondary component's cool core) moves outward. After the two cores reach maximum separation, the secondary core falls back toward the primary core and is accreted. Finally the system evolves into a single merger remnant. 

{Three lines} of evidence suggest that MS\,0451 is in a post-collision state, approximately 2--7\,Gyr after the first core passage of two progenitors that are now approaching second apocentre:
\begin{itemize}
\item While we observe a bimodal (Sect.~\ref{sec:sl}) 
and elongated (Sect.~\ref{Sect:haloshape}) distribution of dark matter, we find a spherical distribution of gas with almost constant central density (Sect.~\ref{sec:ICM}). Such a  contrasting configuration is seen in simulations of major mergers 2--7\,Gyr after first core-core interaction,``following the merger, the resultant system settles into virial equilibrium sooner than into hydrostatic equilibrium'' \citep{Poole:2006nm}.
This period, in which the gas has not yet had time to fully relax and settle into the gravitational potential of the combined halo, represents the second infall phase before the system's final relaxation. These findings apply to a wide range of initial conditions regarding the progenitor mass ratio ($M_{\rm{primary}}:M_{\rm{secondary}}=$1:1, 3:1, 10:1) and the ratio between the secondary's transverse velocity  and the primary's circular velocity, a quantity that affects impact parameters ($v_t/v_c=0.0, 0.15, 0.45$).

\item In simulations, merging increases the entropy of gas in the cluster core, leading to a large core radius and low concentration \citep{Ritchie:2001qt}, exactly as seen in MS\,0451 (Fig.~\ref{fig:gas_entropy}).
\item The cluster is connected to its large-scale environment through a number of substructures (Sect.~\ref{sc:substructures}) and possible filaments (Sect.~\ref{sc:filament}).  The distinctly asymmetric distribution of these features differs starkly from that of simulated clusters which grow through smooth, continuous accretion without strong directional preferences, and whose major axes are aligned with filaments in opposite directions.
\end{itemize}

{The dynamical history inferred from our analysis of MS\,0451 provides a possible explanation for the star-formation history observed in this system. In a comparative study of massive clusters, \cite{2007ApJ...665.1067M,2007ApJ...671.1503M} used optical and near-UV spectroscopy of passive spirals within 1.5\,Mpc of the cluster core to conclude that the star formation in MS\,0451 was abruptly quenched at a redshift of $z=2$, i.e., $\sim$5\,Gyr before the redshift of observation, consistent with our estimate of the temporal evolution of the merger event. \cite{2007ApJ...671.1503M} ascribe the sudden cessation of star formation to ram-pressure stripping by a particularly dense ICM, as evidenced by MS\,0451's bright, extended X-ray emission observed today. The merger scenario proposed by us lends strong support to this explanation, by adding a contemporaneous second component of intra-cluster gas, moving at high relative velocity through the cluster core and thus dramatically increasing the ram pressure \citep[c.f.][]{1999PASJ...51L...1F,2008A&A...481..337K,2013MNRAS.435.2713V}. Beyond 1.5\,Mpc, the passive spirals in MS\,0451 show a ``starvation-like'' gradual cessation of star formation, consistent with secular pre-processing in infalling groups. From their classification of galaxy morphologies in the {\it HST} imaging data, \cite{2007ApJ...671.1503M} also concluded that passive spirals are all but absent inside the central 600\,kpc, having evolved into S0 galaxies. 
This finding too could be the result of ongoing, enhanced ram-pressure stripping by the current ICM as galaxies fall towards the cluster core. Alternatively, the lack of passive spirals within the inner regions of MS\,0451 could be a residual indication of the merging subhalo's trajectory, as it takes $\sim400$\,Myr for a galaxy to travel $\sim$600\,kpc across MS\,0451. Our inferences about MS\,0451's merger dynamics thus complement and support the conclusions of previous studies of its star-formation history.}

\section{Conclusions}
\label{sec:concs}
We present the first combined strong- and weak-gravitational lensing analysis of the massive galaxy cluster MS\,0451, exploiting the largest mosaic of \emph{HST} imaging around any massive cluster: 41 ACS pointings covering an area of $\sim$20$\times$20\,arcmin$^{2}$ ($\sim$6$\times$6\,Mpc$^{2}$). The strong-lensing model exploits 16 multiple-image systems, and our weak-lensing analysis uses a catalogue of 20,138 background galaxies ($\sim$44 arcmin$^{-2}$). We combine these constraints using the \textsc{Lenstool} multi-scale grid technique.

The reconstructed mass distribution of MS\,0451 reveals a bimodal cluster core, elongated along the South-East to North-West axis and surrounded by six substructures, as well as eight weak-lensing peaks created by mass concentrations at other redshifts projected along our line of sight. {We find a total mass of the system of $M_{200}=(1.65\pm0.24)\times10^{15}\,\msun$ with an NFW concentration of $c_{200}=3.79\pm0.36$ (the gNFW and Einasto models yield similar results, while a Burkert model is disfavoured).} Our mass map is consistent with that of the most recent ground-based weak-lensing analysis \citep{2015A&A...581A..31S} but resolves three times more substructures at equivalent significance of detection. The mass distribution of MS\,0451 becomes more circular at large radii, parameterized by an axis ratio that increases from $q=b/a=0.48\pm0.01$ within a projected radius of $R=640$\,kpc to $q=0.57\pm0.03$ inside $R=3.2$\,Mpc. A flattening of MS\,0451's density profile at $R\approx 2$\,Mpc is well fit by the splashback feature in the DK14 model. However, this model's additional complexity negates the  improved fit according to Bayesian Information Criteria; the aforementioned flattening may thus just be noise or due to large-scale structure projected from other redshifts.

In our X-ray analysis, we measure a baryonic-mass fraction of $f_{\rm b,500}=(14.6\pm1.4)\%$ for MS\,0451, consistent with the cosmic baryon fraction \citep{2016A&A...594A..13P}, and a total mass of $M_{200}=(1.75\pm 0.75)\times10^{15}\,\msun$, in good agreement with the lensing estimate. We note though that the assumption of hydrostatic equilibrium underlying the X-ray mass measurement is unlikely to be valid, given that the cluster's gas is distributed very differently than the dark matter. The distribution of gas is circularly symmetric, with a constant-density core and low concentration, $c_{200,\rm{HSE}}=2.35^{+0.89}_{-0.70}$. We also find a strong excess of gas entropy in the cluster's central 300\,kpc.

{Similarly contrasting distributions of gas and dark matter are seen in simulations of post-merger clusters \citep{Poole:2006nm}. The matter distribution observed in MS\,0451 suggests that the cluster underwent a major merger $\sim$2--7\,Gyr ago, and that the two dark-matter halos in the centre are now approaching second apocentre. This merger would have quenched star formation, as ram-pressure from the dense ICM stripped cold gas from cluster member galaxies. Thus the evidence from gravitational lensing, X-ray emission, optical photometry, and spectroscopy all point to a consistent dynamical history.}

We find tentative evidence of three filaments extending from the cluster. The distribution of substructures and a noisy measurement of aperture multipole moments indicate that all three point in similar directions, between East and South. Interestingly, their distribution is asymmetric, with no counterparts to the North or West. Aperture multipole moments appear to be a promising method to detect extended filaments. However, our measurements based on single-orbit \emph{HST} data are dominated by shape noise, and deeper space-based observations will be necessary to robustly test this method.

In the next decade, wide-field, space-based surveys at high resolution are planned as part of the \emph{Euclid} and the \emph{Nancy Grace Roman Space Telescope} missions, as the Vera C. Rubin Observatory becomes operational on the ground. MS\,0451 will be an ideal target for future studies to characterise infalling substructures along filaments, the timing of star-formation processes after a major merger, and the late-time evolution of cluster collisions.

\section*{Acknowledgements}
We would like to thank the referee for giving useful comments and improving our manuscript. We are grateful for constructive discussions of this work with A.\ Edge, A.\ Niemiec, and J.\ Nightingale. We thank N.\ Martinet for sharing unpublished results from his analysis, allowing a comparison between our two studies. SIT is supported by Van Mildert College Trust PhD Scholarships. MJ is supported by the United Kingdom Research and Innovation (UKRI) Future Leaders Fellowship `Using Cosmic Beasts to uncover the Nature of Dark Matter' [grant number MR/S017216/1]. RM is supported by a Royal Society University Research Fellowship. DH acknowledges support by the ITP Delta foundation. AR is supported by the European Research Council (ERC-StG-716532-PUNCA) and STFC (ST/N001494/1). 

\emph{Facilities:} This paper uses data from observations GO-9836 with the NASA/ESA
\emph{Hubble Space Telescope}, obtained at the Space Telescope Science Institute, which is operated by AURA Inc, under NASA contract NAS 5-26555. All data are available from the telescopes' archives. This paper uses data from observations (ID: 0205670101) obtained with \emph{XMM-Newton}, an ESA science mission with instruments and contributions directly funded by ESA Member States and NASA. This paper uses data from observations (ID: 06BH34, PI: H.\ Ebeling) obtained with MegaPrime/MegaCam, a joint project of Canada-France-Hawaii Telescope (CFHT) and CEA/DAPNIA, at the CFHT which is operated by the National Research Council (NRC) of Canada, the Institut National des Sciences de l'Univers of the Centre National de la Recherche Scientifique (CNRS) in France, and the University of Hawaii. This work is based in part on data products produced at \textsc{terapix}, and the Canadian Astronomy Data Centre as part of the CFHT Legacy Survey, a collaborative project of NRC and CNRS. This paper uses data from observations (ID: 08BH63, 07BH98, PI: C.-J.\ Ma) obtained with WIRCam, a joint project of CFHT, the Academia Sinica Institute of Astronomy and Astrophysics (ASIAA) in Taiwan, the Korea Astronomy and Space Science Institute (KASI) in Korea, Canada, France, and the Canada-France-Hawaii Telescope (CFHT) which is operated by the National Research Council (NRC) of Canada, the Institut National des Sciences de l'Univers of the Centre National de la Recherche Scientifique of France, and the University of Hawaii. This work is also based on data collected at Subaru Telescope, which is operated by the National Astronomical Observatory of Japan. This project was also supported by the Science and Technology Facilities Council [grant number  ST/P000541/1].
This work used the DiRAC Durham supercomputing facility, which is managed by the Institute for Computational Cosmology on behalf of the STFC DiRAC HPC Facility (www.dirac.ac.uk). Its equipment was funded by BEIS capital funding via STFC capital grants ST/K00042X/1, ST/P002293/1 and ST/R002371/1, Durham University and STFC operations grant ST/R000832/1. DiRAC is part of the UK National e-Infrastructure.

\section*{Data Availability}
The data underlying this article are available from the corresponding author on reasonable request.

\bibliographystyle{mnras}
\bibliography{paper3}

\appendix

\section{$\!\!$Mass mapping with KS93 and MRLens}
\label{sec:massrec_mrlens}
As an alternative to the \textsc{Lenstool} multiscale grid method (Sect.~\ref{sec:massrec}), we also explore the \citet{1993ApJ...404..441K} direct-inversion method to map the lensing mass. Direct inversion converts an observed, binned shear map, $\boldsymbol\gamma(\boldsymbol{R})$, into a convergence map, $\kappa(\boldsymbol{R})$, via their Fourier transforms $\hat{\boldsymbol\gamma}(\boldsymbol{k})$ and $\hat{\kappa}(\boldsymbol{k})$:
\begin{equation}
\hat\kappa(\boldsymbol{k})=
\frac{1}{2}\left(\frac{k_1^2-k_2^2}{k_1^2+k_2^2}\right)\hat\gamma_1(\boldsymbol{k})+
\frac{1}{2}\left(\frac{k_1k_2}{k_1^2+k_2^2}\right)\hat\gamma_2(\boldsymbol{k})\, ,
\end{equation}
where $\boldsymbol{k} = (k_1, k_2)$ is the wave-vector conjugate to $\boldsymbol{R}$. We implement this method using a top-hat window function of radius 
$R_f=0.2\arcmin$, $S(\Delta R) = H(R_f-|\Delta R|)$, where $H(x)$ is the Heaviside function, \citep{Merten:2008qf,Umetsu:2015hda} to (re)bin the weak-lensing shear catalogues $\boldsymbol{\gamma}$ into $340\times340$\,pixel grids; we also use discrete Fourier transforms. For the region outside the {\emph{HST}} fied of view, we mitigate boundary effects by zero padding out to 105\arcmin$\times$105\arcmin. Note that this method does not make use of any strong-lensing information.

To denoise the convergence map, we apply the \textsc{MRLens} (Multi-Resolution methods for gravitational Lensing) software\footnote{We implement \textsc{MRLens} using the June 26, 2017 version of software available from \url{https://www.cosmostat.org/software/mrlens}.}. The noisy 2D convergence map is  decomposed into six wavelet scales and filtered in ten iterations. We refer the reader to \cite{2006A&A...451.1139S} for a complete description of the \textsc{MRLens} algorithm.

\begin{figure}
\centering
\includegraphics[width=0.5\textwidth]{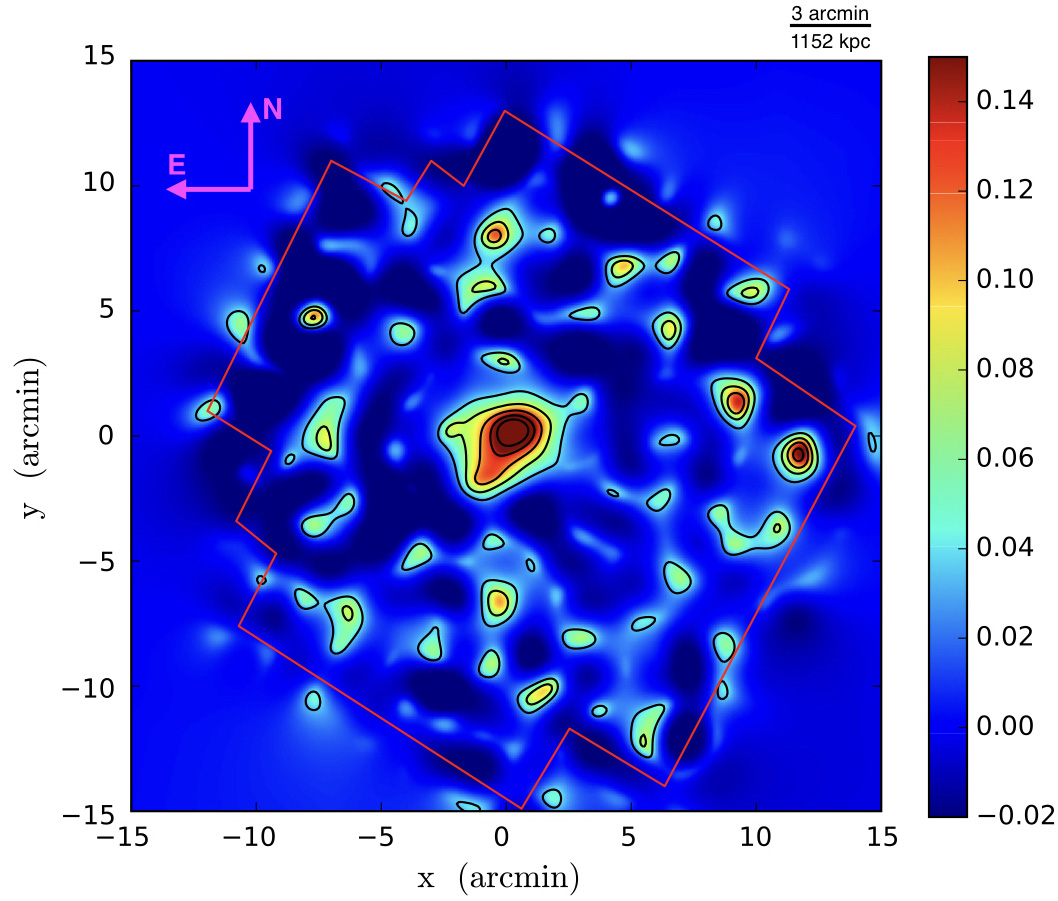}
\caption{Convergence map of MS\,0451 obtained with an alternative method that combines \textsc{KS93} and \textsc{MRLens}.
The field is centred at the location of the BCG, with a red polygon indicating the extent of the \emph{HST} imaging mosaic. Black contours show statistical significance thresholds, starting at 1\,$\sigma_{\kappa}$ and spaced linearly in units of 1\,$\sigma_{\kappa}$.
Overall the map is consistent with the results of our {\sc Lenstool} method (Fig.~\ref{fig:ms0451_massrec}), but noisier and of lower resolution. } 
\label{fig:ms0451_KS93}
\end{figure}

Although the resulting mass map (Fig.~\ref{fig:ms0451_KS93}) is noisier than the \textsc{Lenstool} map (Fig.~\ref{fig:ms0451_massrec}), the cluster core is still detected at more than 4$\sigma_{\kappa}$ significance, where $\sigma_{\kappa}$ is the standard deviation of $\kappa$ over all pixels within the \emph{HST} field of view. The MS\,0451 reconstruction also shows an elongation along the South-East to North-West direction, consistent with other  analyses \citep{2015A&A...581A..31S,2016A&A...590A..69M} that use only weak-lensing data. Several weak-lensing peaks are detected at lower  statistical significance (1--3$\sigma_{\kappa}$) than with \textsc{Lenstool}, including all confirmed substructures.
Overall, the level of noise and spurious peaks  (mainly from line-of-sight projections) in this convergence map are similar to what is found in maps reconstructed from mock observations of simulated clusters 
(see Fig.~3 in \cite{2020arXiv200610155T}).

\section{radial density profile models}
\label{sec:appendix:1dprofiles}

In Section~\ref{sec:1dprofiles}, we fit the observed mass distribution in MS\,0451 using analytical models of the 3D density; from these we compute the projected surface density by integrating along the line of sight,
\begin{equation}
\Sigma(R)=2\int_R^{\infty}\frac{r\rho(r)}{\sqrt{r^2-R^2}}dr.
\end{equation}
We here briefly summarize the five models investigated in our work.

\paragraph*{\textit{NFW profile --}}
Numerical simulations \citep{Navarro:1995iw,Navarro:1996gj} suggest that dark-matter halos can be described by a universal density profile with a two-parameter functional form of
\begin{equation}
\rho_{\rm{NFW}}=\frac{\rho_s}{(r/r_s)(1+(r/r_s))^2}
\end{equation}
where $\rho_s$ and $r_s$ are the characteristic density and radius, respectively. At $r=r_s$, the logarithmic density slope equals the isothermal value, $d\text{ln}\rho(r)/d\text{ln}r=2$. 
For any given cosmology and cluster redshift, this model can also be parametrized in terms of the concentration parameter $c_{200}\equiv R_{200}/r_s$ (here $R_{200}$ is the radius at which the mean density is equal to 200 times the critical density of the Universe, $\rho_{c}$) and the halo mass,  $M_{200}=(4\pi/3) 200\rho_{c}R_{200}^3$.

\paragraph*{\textit{Generalized NFW (gNFW) profile --}}
A generalized version of the NFW model, of the form
\begin{equation}
\rho_{\rm{gNFW}}=\frac{\rho_s}{(r/r_s)^{\alpha}(1+(r/r_s))^{(3-\alpha)}}
\end{equation}
\citep{1996MNRAS.278..488Z}.
This profile features a power-law-shaped central cusp, $\rho\propto r^{-\alpha}$ and reduces to a NFW model for $\alpha$ = 1. Generalizing the approach taken for the  NFW model, we describe the gNFW profile with a central slope $\alpha$, a halo mass $M_{200}$, and a concentration $c_{200}=r_{200}/(2-\alpha)\,r_s$. 
The radial dependence of the gNFW lensing signal was calculated by \cite{Keeton:2001ss}.

\paragraph*{\textit{Einasto profile--}}
Several N-body simulations \citep[e.g.][]{Graham:2005xx,2015Sci...347.1462H,2014MNRAS.441.3359D,Klypin:2014kpa} have shown that cold dark matter (CDM) halos can be best described by the Einasto density profile \citep[][]{1965TrAlm...5...87E}, which is given by 
\begin{equation}
\rho_{\text{Einasto}}=\rho_s \exp{\left\{-\frac{2}{\alpha_E}\left[\left(\frac{r}{r_s}\right)^{\alpha_E}-1\right]\right\}},
\end{equation}
where $\alpha_E$ is the shape parameter describing the steepness of the logarithmic slope. Alternatively, we can express the density profile as
\begin{equation}
\frac{d\log{\rho}}{d\log{r}}=-2\left(\frac{r}{r_s}\right)^{\alpha_E}.
\end{equation}
An Einasto profile with $\alpha_E\sim0.18$ has a similar shape as a NFW profile at a given concentration \citep[][]{Ludlow:2013bd}. 

\paragraph*{\textit{Burkert profile --}}
The  dark-matter halo proposed by \cite{1995ApJ...447L..25B} is described by a density profile 
\begin{equation}
\rho_{\rm{Burkert}}=\frac{\rho_{\rm{core}}}{(1+r/r_{\rm{core}})(1+r^2/r_{\rm{core}}^2)},
\end{equation}
where $\rho_{\rm{core}}$ and $r_{\rm{core}}$ parameterise the density and size of a (constant density) core.

\paragraph*{\textit{Diemer \& Kravtsov profile --}}
The more flexible density profile proposed by \citet[][hereafter DK14]{Diemer:2014xya} was calibrated using a suite of $\Lambda$CDM simulations and  is described by two components: (1) collapsed matter, modelled by a truncated Einasto profile \citep{1965TrAlm...5...87E}, and (2)  infalling material, modelled by a power law function. The complete model is given by
\begin{equation}
\rho(r)=\rho_{\text{Einasto}}(r)\times f_{\text{trans}}(r)+\rho_{\text{infall}}(r)
\end{equation}

\begin{equation}
f_{\text{trans}}(r)=\left[1+\left(\frac{r}{r_t}\right)^\beta\right]^{-\gamma/\beta}
\end{equation}

\begin{equation}
\rho_{\text{infall}}(r)=\frac{\rho_m b_e}{\Delta_{\text{max}}^{-1}+\left(\frac{r}{5r_{200m}}\right)^{s_e}}
\end{equation}
 where $\Delta_{\text{max}}=10^{3}$, and the transition term $f_{\text{trans}}$ captures the steepening of the profile around a truncation radius $r_t$. The shape parameters $\gamma$ and $\beta$ determine the steepness of the profile 
 and how quickly the slope changes, respectively. For the infalling material, the power-law profile (which decreases with radius, since $s_e$>0) approaches the mean density of the Universe, $\rho_m$, at large radii. $\Delta_{\text{max}}$ is introduced to avoid a spurious contribution toward the center of the cluster. We adapted the publicly available code \textsc{COLOSSUS} \citep{Diemer:2017bwl} to calculate the DK14 density profile.

\section{Splashback radius}
\label{sec:appendix:splashback}

\begin{table}
\centering
\begin{tabular}{cr@{$\,\pm\,$}l}
\hline
\hline
\textbf{Parameter} & \multicolumn{2}{c}{\textbf{Constraint}} \\
\hline
$\rho_s\,[10^{3}\msun/\rm{kpc}^{3}]$ & $598.32$ & $6.41$  \\
$r_s\,[\rm{Mpc}]$ &  $0.33$ & $0.12$  \\
$r_t\,[\rm{Mpc}]$ & $1.76$ & $1.34$\\
$\log{\alpha}$ &  $-0.78$ & $0.39$ \\
$\log{\beta}$ &  $0.87$ & $0.45$\\
$\log{\gamma}$ &  $0.70$ & $0.45$\\
$b_{e}$ & $1.96$ & $1.00$\\ 
$s_{e}$ & $1.75$ & $0.54$\\ 
\hline
\hline
\end{tabular}
\caption{Marginalized posterior constraints on the DK14 model.}
\label{tab:DK14tab}
\end{table}

\begin{figure}
\centering
\includegraphics[width=0.5\textwidth]{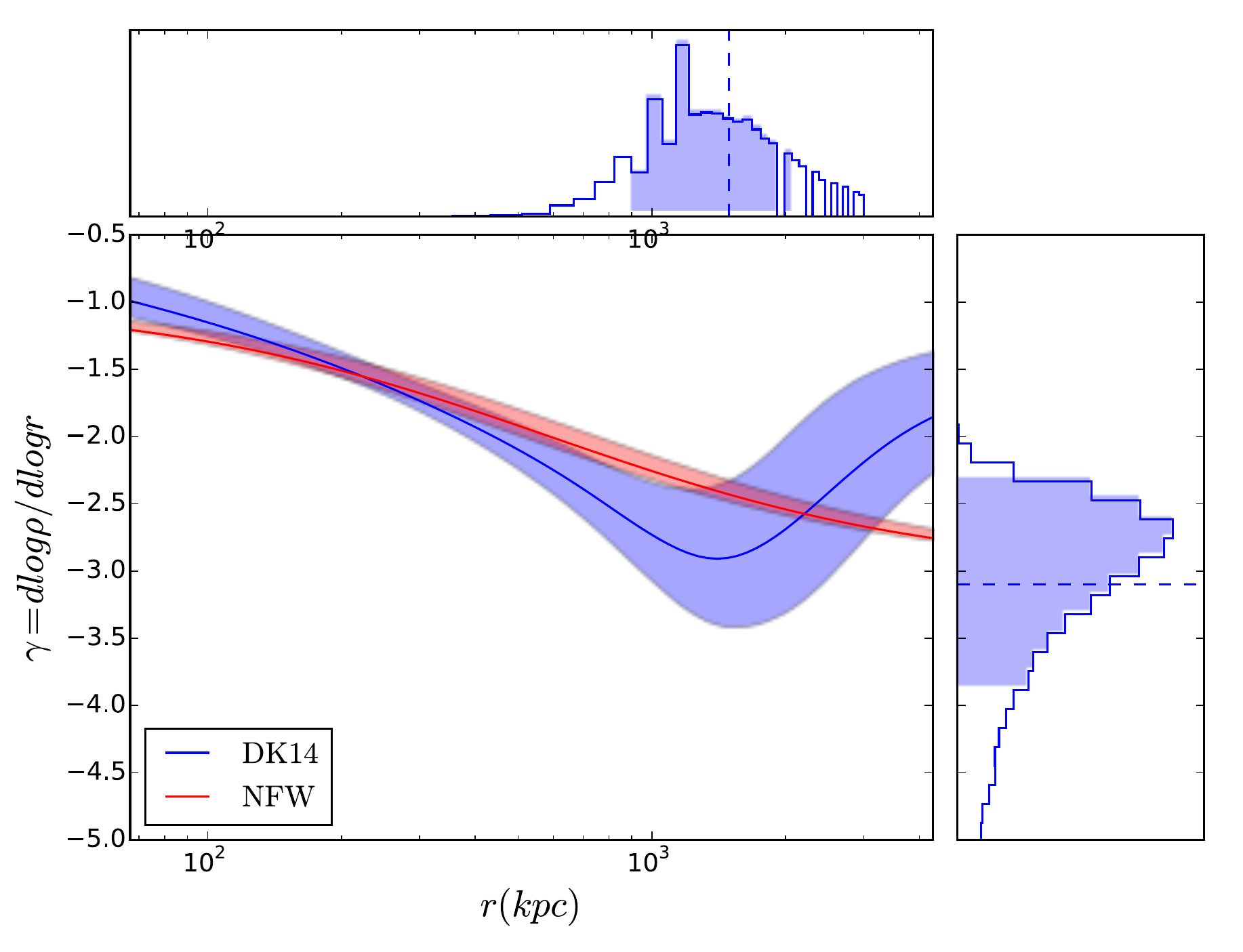}
\caption{
Radial gradient of the total mass distribution of MS\,0451 from fitted NFW (red) and DK14 (blue) models. Solid lines show the mean inferred values; shaded regions show 68\% confidence intervals 
The upper and right panels show the posterior probability distributions of the splashback radius $r_{\rm{sp}}$ and the gradient at the splashback radius $\gamma(r_{\rm{sp}})$. 
Dashed lines and shaded regions indicate the mean and 68\% confidence intervals respectively.} 
\label{fig:splashbackr}
\end{figure}

$N$-body simulations by \cite{Diemer:2014xya} show that density profiles of dark-matter halos exhibit a sharp steepening close to the virial radius. This feature depends on the accretion history of the cluster  and results from an absence of
particles orbiting beyond the radius of second turnaround. It gives us a  physically motivated definition for the boundary of dark-matter halos. Here we investigate the splashback feature of MS\,0451 by fitting its density profile with a DK14 profile (Sect.~\ref{Sec:densityprofile}). The marginalized posterior constraints are listed in Table~\ref{tab:DK14tab} where we employ the biweight estimators of \cite{1990AJ....100...32B} for
the center and dispersion of the marginalized posterior
distributions \citep[e.g.][]{2011MNRAS.416.3187S,Umetsu:2014vna,Chiu:2018gok}.

Following \cite{2015ApJ...810...36M} we define the splashback radius $\rm{r_{\rm{sp}}}$ as the location of a local minimum in the logarithmic slope of the density profile, $\gamma\equiv\rm{d\,log}\rho/\rm{d\,log}r$. 
Figure~\ref{fig:splashbackr} shows the mean and 68\% confidence intervals of $\gamma$ inferred from the DK14 fit, together with the posterior probability distribution of the splashback radius, and the posterior probability distribution of $\gamma(r_{\rm{sp}})$. 
The biweight central location of $\gamma(r_{\rm{sp}})$ is $-3.10\pm0.74$
at $r_{\rm{sp}} =(1.49\pm0.57)$\,Mpc. 
This is not significantly different from the value of  $\gamma\sim -2.4$ for the best-fit NFW model at this radius, and the Bayesian and Akaike Information Criteria disfavour  the DK14 model because of its increased complexity.
Furthermore, our best-fit $r_{\rm{sp}}$ value is lower than predictions from cosmological simulations and other observational analysis \citep[e.g.][]{Contigiani:2018qxn}.

It is possible that a true splashback feature lies close to (or ouside) the edge of the \emph{HST} field of view, where `noise' in the form of lensing signal from projected substructures exceeds the lensing signal of the cluster and is correlated between radial bins.
Hence, similarly to the findings obtained by \cite{Umetsu:2016cun} for a sample of 16 clusters, our measurements for MS\,0451 place only  a lower limit on the splashback radius, i.e., $r_{\rm{sp}}>1.49$\,Mpc.

\section{Redshift distribution of detected weak-lensing peaks}
\label{sec:appendix:substructurez}

\begin{figure*}
\centering
\includegraphics[width=0.9\textwidth]{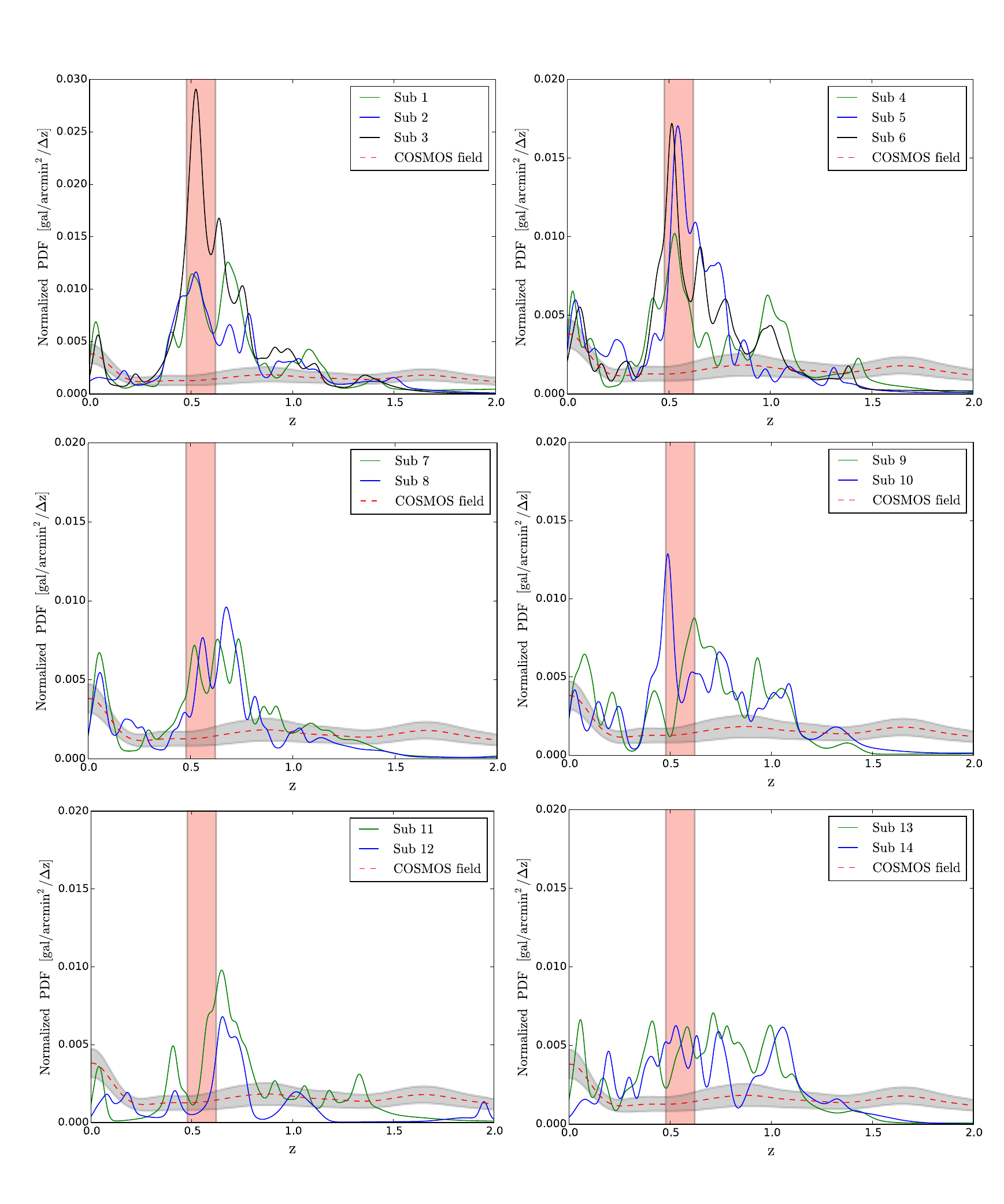}
\caption{Normalized PDF of photometric redshifts $z_\mathrm{phot}$ for all galaxies within a circular aperture ($R=480$\,kpc) of each of the 14 detected weak-lensing peaks. 
Within the typical  uncertainties of photometric redshifts, the galaxy overdensities at $0.48<z_\mathrm{phot}<0.61$ (vertical red bands) are consistent with being at the redshift of MS\,0451. 
For comparison, the red dashed line shows the redshift distribution of galaxies detected in comparable {\emph HST} imaging of a blank patch of sky (the COSMOS field) and the associated $1\sigma$ scatter (grey band).\label{lastpage}
} 
\label{fig:zdistri_app}
\end{figure*}

The summed probability density functions (PDFs) of photometric redshifts $z_\mathrm{phot}$ for all galaxies within $R=480$\,kpc of each substructure are shown in Fig.~\ref{fig:zdistri_app}. For comparison, we also show the redshift distribution of galaxies observed in {\emph HST} imaging in a blank region of sky (the COSMOS field), within the same passband and to the same depth,  normalised to the number density of galaxies in our catalogue that have photometric redshifts (a higher fraction of COSMOS galaxies have photometric redshifts, particularly at high redshift, thanks to the deeper Subaru imaging). The $1\sigma$ scatter in the redshift distribution of COSMOS galaxies, calculated using the bootstrap method, is unchanged by the higher precision of $z_\mathrm{phot}$ measurements that use many more colours.

Substructures 1, 2, 3, 4, 5 and~6 are dominated by galaxies at the same redshift as the main cluster, and are therefore likely to be physically associated. 
By contrast, Substructures 7 to 14 (marked by black circles in Fig.~\ref{fig:MS0451_subs}) are dominated by galaxies at different redshifts, or at a mixture of redshifts, and are therefore implausible as parts of MS\,0451. We note that Substructures 8, 9 and 11 might be a linked system behind the cluster ($z\sim0.7$) and even appear as an extended mass distribution in the AMM map (Fig.~\ref{fig:Q_MS0451}).

\bsp
\end{document}